\documentclass[draftcls,peerreview,12pt,onecolumn]{IEEEtran}
\usepackage{amsmath}
\usepackage[dvips]{graphicx}
\usepackage{amssymb}
\usepackage{cite}

\title{A Comparative Study of Relaying Schemes with Decode-and-Forward over Nakagami-$m$ Fading Channels}

\author{George~C.~Alexandropoulos, Agisilaos~Papadogiannis and Paschalis~C.~Sofotasios
\thanks{G. C. Alexandropoulos is with the Department of Telecommunications Science and Technology, University of Peloponnese, End of Karaiskaki Street, GR-22100 Tripolis, Greece  (e-mail: alexandg@ieee.org).}
\thanks{A. Papadogiannis is with the Communications Research Group, Department of Electronics, University of York, York, YO10 5DD, United Kingdom (e-mail: ap851@ohm.york.ac.uk).}
\thanks{P. C. Sofotasios is with School of Electronic and Electrical Engineering, University of Leeds, Leeds, LS2 9JT, United Kingdom (e-mail: p.sofotasios@leeds.ac.uk).}
}

\begin{document}

\maketitle
\markboth{Submitted to \textit{Journal of Computer Systems, Networks, and Communications}}{}

\begin{abstract}
Utilizing relaying techniques to improve performance of wireless systems is a promising avenue. However, it is crucial to understand what type of relaying schemes should be used for achieving different performance objectives under realistic fading conditions. In this paper, we present a general framework for modelling and evaluating the performance of relaying schemes based on the decode-and-forward (DF) protocol over independent and not necessarily identically distributed (INID) Nakagami-$m$ fading channels. In particular, we present closed-form expressions for the statistics of the instantaneous output signal-to-noise ratio of four significant relaying schemes with DF; two based on repetitive transmission and the other two based on relay selection (RS). These expressions are then used to obtain closed-form expressions for the outage probability and the average symbol error probability for several modulations of all considered relaying schemes over INID Nakagami-$m$ fading. Importantly, it is shown that when the channel state information for RS is perfect, RS-based transmission schemes always outperform repetitive ones. Furthermore, when the direct link between the source and the destination nodes is sufficiently strong, relaying may not result in any gains and in this case it should be switched-off.
\end{abstract}

\newpage
\begin{keywords}
Bit error probability, cooperative diversity, Nakagami-$m$ fading, outage probability, relay selection, repetitive transmission.
\end{keywords}

\section{Introduction}
The significance of multiple-input multiple-output (MIMO) techniques for modern wireless systems has been well appreciated. Multiple collocated antennas can improve transmission reliability and the achievable capacity through diversity, spatial multiplexing and/or interference suppression \cite{J:Foschini_2010,B:Biglieri_2007}. However, the cost of mobile devices is proportional to their number of antennas and this creates a serious practical limitation for the use of MIMO. Cooperative diversity is a promising new avenue which allows cooperation amongst a number of wireless nodes which effectively profit from MIMO techniques without requiring multiple collocated antennas \cite{J:Laneman_2003,J:Sendonaris_2003}.

Dual-hop cooperative diversity entails that the transmission of the source node towards a destination node is assisted by one or more relay nodes which can be seen to form a conceptual MIMO array \cite{J:Laneman_2003}. Relay nodes can either be fixed, being part of the system infrastructure, or mobile, i.e., mobile nodes that relay signals intended for other mobile nodes \cite{J:AgisilaosGeorge_2011}. Cooperative diversity is shown to improve the transmission reliability and the achievable capacity while it also extends coverage. Essentially, it can achieve diversity gains and has the additional advantage over conventional MIMO that the remote cooperating antennas experience very low or inexisting correlation \cite{J:Laneman_2003,J:Bletsas_2006,J:Agisilaos_ASIMOLAR_08,C:Agisilaos_GC_2009}. The performance of a system exploiting cooperative diversity depends on the employed relaying protocol and scheme, i.e., the way of utilizing relay nodes \cite{J:Laneman_2003,J:Bletsas_2006,J:Agisilaos_ASIMOLAR_08,C:Agisilaos_GC_2009}. Consequently, it is crucial to gain insights on which relay scheme is most suitable for achieving a particular objective; some common objectives are the minimization of the outage probability (OP) or the average symbol error probability (ASEP) \cite{J:Yan_OP_Nak_2009,J:Geo_Agi_RS_2010}.

In the present work, we consider the decode-and-forward (DF) relaying protocol under the assumption that the message transmitted by the source is decoded and retransmitted to destination by one or more relays in a dual-hop fashion. We also take into account four relaying schemes; two based on repetitive transmission and the other two based on relay selection (RS). According to repetitive transmission, all relays that decode source's message retransmit it repetitively to the destination node which employs diversity techniques to combine the different signal copies \cite{J:Beaulieu_DF_2006,J:Hu_DF_2007}. One version of RS entails that amongst relays decoding the source's message only one is selected to retransmit it to destination, the one with the strongest relay to destination channel \cite{J:Bletsas_2006,J:Beres_2008}. Another version of RS utilizes the best relay only in the case that it results in capacity gains over the direct source to destination transmission \cite{J:Agisilaos_ASIMOLAR_08,J:Geo_Agi_RS_2010}. In the literature such schemes have been considered partially and mainly assuming independent and identically distributed (IID) Rayleigh fading channels \cite{J:Beaulieu_DF_2006,J:Bletsas_2006,J:Beres_2008}. Recently, the Nakagami-$m$ fading model has received a lot of attention as it can describe more accurately the fading process and helps in understanding the gains of cooperative diversity \cite{J:Xu_2009, J:Yan_OP_Nak_2009, J:Datsikas_2008, J:Agi_Geo_PIMRC09, J:Geo_Agi_RS_2010, J:Geo_Agi_PIMRC10, J:Trung_Geo_TVT11}. However, there has not been a complete study that sheds light on the question of which relaying scheme is preferable and under which channel conditions.

In this paper, we present a general analytical framework for modeling and evaluating performance of relaying schemes with DF under independent and not necessarily identically distributed (INID) Nakagami-$m$ fading channels. Further to this, we obtain closed-form expressions for the OP and the ASEP performance of the RS and repetitive schemes when maximal ratio diversity (MRD) or selection diversity (SD) are employed at the destination node. We conclude that the RS-based transmission always performs better in terms of OP and ASEP than repetitive transmission when channel state information (CSI) for RS is perfect. In addition, when the direct source to destination link is sufficiently strong, relaying should be disabled when the objective is the minimization of OP. Although RS requires only two time slots for transmission (the repetitive scheme needs as many time slots as the number of decoding relays) its performance heavily relies on the quality of CSI for RS.

The remainder of this paper is structured as follows: Section~\ref{Sec:System} outlines the system and channel models. Section~\ref{Sec:MGF} presents closed-form expressions for the statistics of the instantaneous output SNR of the considered DF relaying schemes over INID Nakagami-$m$ fading channels. In Section~\ref{Sec:Performance}, closed-from expressions are derived for the OP and ASEP performance of all relaying schemes. Section~\ref{Sec:num_res} contains numerical results and relevant discussion, whereas Section~\ref{Sec:concl} concludes the paper.

\textit{Notations:} Throughout this paper, $\left|\mathbb{A}\right|$ represents the cardinality of the set $\mathbb{A}$ and ${\mathbb E}\langle\cdot\rangle$ denotes the expectation operator. $\Pr\left\{\cdot\right\}$ denotes probability, $\mathbb{L}^{-1}\{\cdot;\cdot\}$ denotes the inverse Laplace transform and $X\sim\mathcal{C}\mathcal{N}\left(\mu,\sigma^{2}\right)$ represents a random variable (RV) following the complex normal distribution with mean $\mu$ and variance $\sigma^{2}$. $\Gamma\left(\cdot\right)$ is the Gamma function \cite[eq.
(8.310/1)]{B:Gra_Ryz_Book} and $\mathsf{\Gamma}\left(\cdot,\cdot\right)$ is the lower incomplete Gamma function \cite[eq. (8.350/1)]{B:Gra_Ryz_Book}. Moreover, $\delta\left(\cdot\right)$ is the Dirac function, $u\left(\cdot\right)$ is the unit step function and $\delta\left(\cdot,\cdot\right)$ is the Kronecker Delta function.

\section{System and Channel Model}\label{Sec:System}
We consider a dual-hop cooperative wireless system, as illustrated in Fig.~\ref{Fig:System}, consisting of $L+2$ wireless nodes: one source node $S$, a set $\mathbb{A}$ of $L$ relay nodes each denoted by $R_k$, $k=1,2,\ldots,L$, and one destination node $D$. All $R_k$'s are assumed to operate in half-duplex mode, i.e., they cannot transmit and receive simultaneously, and node $D$ is assumed to possess perfectly $S \rightarrow D$ and all $R_{k} \rightarrow D$ CSI. We consider orthogonal DF (ODF) relaying \cite{J:Laneman_2003} for which each $R_k$ that successfully decodes $S$'s signal retransmits it to $D$; during each $R_k$'s transmission to $D$ node $S$ remains silent. Assuming repetitive transmission \cite{J:Laneman_2003}, $L+1$ time slots are used to forward $S$'s signal to $D$ in a predetermined order, whereas only two time slots are needed with RS-based transmission \cite{J:Bletsas_2006}. In particular, during the first time slot for both transmission strategies, $S$ broadcasts its signal to all $R_k$'s and also to $D$. Considering quasi-static fading channels, the received signal at $R_k$'s and $D$, respectively, during the first time slot can be mathematically expressed as
\begin{subequations}\label{eq:Received_FirstSlot}
\begin{equation}\label{Eq:YD1}
y_D^{(1)} = h_{SD}s+n_{SD}
\end{equation}
\begin{equation}\label{Eq:YR}
y_{R_k} = h_{SR_k}s+n_{SR_k}
\end{equation}
\end{subequations}
where $h_{SD}$ and $h_{SR_k}$ denote the $S\rightarrow D$ and $S\rightarrow R_k$, respectively, complex-valued channel coefficients and $s$ is the transmitted complex message symbol with average symbol energy $E_s$. Moreover, the notations $n_{SD}$ and $n_{SR_k}$ in \eqref{eq:Received_FirstSlot} represent the additive white Gaussian noise (AWGN) of the $S\rightarrow D$ and $S\rightarrow R_k$ channel, respectively, with $n_{SD},n_{SR_k}\sim\mathcal{C}\mathcal{N}\left(0,N_0\right)$. For both $n_{SD}$ and $n_{SR_k}$, it is assumed that they are statistically independent of $s$.

Let us assume that set $\mathbb{B}\subseteq\mathbb{A}$ contains the relay nodes that have successfully decoded $S$'s signal during the first time slot of transmission. When repetitive transmission is utilized, $L$ more time slots are used for $R_k$'s belonging to $\mathbb{B}$ to forward $s$ to $D$; each $R_k$ retransmits $s$ during the $k$-th time slot\footnote{Note that the assignment of each time slot to $R_k$'s is performed in a predetermined order \cite{J:Laneman_2003}. Thus, due to different $R_k\rightarrow D$ $\forall\,k$ channel conditions and ODF relaying, there might be some unused time slots.}. Hence, for quasi-static fading, the received signal at $D$ at the $k$-th time slot can be expressed as
\begin{equation}\label{Eq:YD_Rep}
y_D^{(k)} = h_{R_kD}s+n_{R_kD}
\end{equation}
with $h_{R_kD}$ representing the $R_k\rightarrow D$ complex-valued channel coefficient and $n_{R_kD}\sim\mathcal{C}\mathcal{N}\left(0,N_0\right)$ is AWGN of this channel that is assumed statistically independent of $s$.

When RS-based transmission is used, one time slot is needed for the relay node $R_{\rm best}$ with the most favourable $R_{\rm best}\rightarrow D$ channel conditions to forward $S$'s signal to $D$. Thus, for this transmission strategy and during the second time slot, the received signal at $D$ for quasi-static fading can be expressed as
\begin{equation}\label{Eq:YD_RS}
y_D^{(2)} = h_{R_{\rm best}D}s+n_{R_{\rm best}D}
\end{equation}
where $h_{R_{\rm best}D}$ and $n_{R_{\rm best}D}\sim\mathcal{C}\mathcal{N}\left(0,N_0\right)$ denote the $R_k\rightarrow D$ complex-valued channel coefficient and the AWGN for this channel, respectively. As \eqref{eq:Received_FirstSlot} and \eqref{Eq:YD_Rep}, it is assumed that $n_{R_{\rm best}D}$ is statistically independent of $s$.

The quasi-static fading channels $h_{SD}$, $h_{SR_k}$ and $h_{R_kD}$ $\forall\,k$, and $h_{R_{\rm best}D}$ are assumed to be modeled as INID Nakagami-$m$ RVs \cite{C:Nakagami}. Let $\gamma_{0} = \left|h_{SD}\right|^2E_\text{s}/N_0$ and $\gamma_{k} = \left|h_{R_{k}D}\right|^2 E_\text{s}/N_0$, $k=1,2,\ldots,L$, be the instantaneous received SNRs of the $S\rightarrow D$ and $R_k\rightarrow D$ link, respectively, with corresponding average values given by $\overline{\gamma}_{0} = \mathbb{E}\left|\langle h_{SD}\right|^2\rangle E_s/N_0$ and $\overline{\gamma}_{k} = \mathbb{E}\langle \left|h_{R_{k}D}\right|^2\rangle E_s/N_0$, respectively. Clearly, each $\gamma_{\ell}$, $\ell=0,1,\ldots,L$, is gamma distributed with probability density function (PDF) given by \cite[Table 2.2]{B:Sim_Alou_Book}
\begin{equation}\label{Eq:Gamma_PDF}
f_{\gamma_{\ell}}\left(x\right) = \frac{C_{\ell}^{m_{\ell}}}{\Gamma\left(m_\ell\right)}x^{m_{\ell}-1}\exp\left(-C_{\ell}x\right)
\end{equation}
where $m_\ell \geq 1/2$ denotes the Nakagami-$m$ fading parameter and $C_{\ell}= m_{\ell}/ \overline{\gamma}_{\ell}$. Integrating \eqref{Eq:Gamma_PDF}, the cumulative distribution function (CDF) of each $\gamma_{\ell}$ is easily obtained as
\begin{equation}\label{Eq:Gamma_CDF}
F_{\gamma_{\ell}}\left(x\right) = \frac{\mathsf{\Gamma}\left(m_\ell,C_{\ell}x\right)}{\Gamma\left(m_\ell\right)}.
\end{equation}
The PDFs and CDFs of the instantaneous received SNRs of the first hop, $\gamma_{L+k} = \left|h_{SR_{k}}\right|^2 E_\text{s}/N_0$ $\forall\,k$, are given using \eqref{Eq:Gamma_PDF} and \eqref{Eq:Gamma_CDF} by $f_{\gamma_{L+k}}\left(x\right)$ and $F_{\gamma_{L+k}}\left(x\right)$, respectively, with fading parameters and average SNRs denoted by $m_{L+k}$ and $\overline{\gamma}_{L+k} = \mathbb{E}\langle \left|h_{SR_{k}}\right|^2\rangle E_s/N_0$, respectively.

\section{Statistics of ODF Relaying Schemes}\label{Sec:MGF}
Relay station nodes that are able to decode the transmitted signal from $S$ constitute the decoding set $\mathbb{B}$. Based on \cite{J:Laneman_2003} and \cite{J:Beres_2008}, for both transmission strategies the elements of $\mathbb{B}$ are obtained as
\begin{equation}\label{Eq:Set_B}
\mathbb{B} = \left\{R_k\in\mathbb{A}:\log_2\left(1+\gamma_{L+k}\right)\geq \alpha\mathcal{R}\right\}
\end{equation}
where $\mathcal{R}$ is $S$'s transmit rate and $\alpha=L+1$ for repetitive transmission, whereas $\alpha=2$ for RS-based transmission. Hence, the probability that $R_k$ does not belong to $\mathbb{B}$ is easily obtained as $\mathcal{P}_k=\text{Pr}\left[\gamma_{L+k}<2^{\alpha\mathcal{R}}-1\right]$. Substituting $F_{\gamma_{L+k}} \left(x\right)$ after using \eqref{Eq:Gamma_CDF}, a closed-form expression for $\mathcal{P}_k$ in INID Nakagami-$m$ fading is given by
\begin{equation}\label{Eq:Prob_k}
\mathcal{P}_k=\frac{\mathsf{\Gamma}\left[m_{L+k},C_{L+k}\left(2^{\alpha\mathcal{R}}-1\right)\right]}{\Gamma\left(m_{L+k}\right)}.
\end{equation}

To analyze the performance of ODF relaying schemes, the $S \longrightarrow D$ direct channel plus the $S\longrightarrow R_k \longrightarrow D$ $\forall\,k$ RS-assisted channels are effectively considered as $L+1$ paths between $S$ and $D$ \cite{J:Beaulieu_DF_2006, J:Datsikas_2008}. In particular, let the zero-th path represent the $S \longrightarrow D$ direct link and the $k$-th path the $S\longrightarrow R_k \longrightarrow D$ cascaded link. We define the instantaneous received SNRs at $D$ of these paths as $g_0$ and $g_k$, respectively. By substituting \eqref{Eq:Gamma_PDF} into \cite[eq. (4)]{J:Beaulieu_DF_2006}, the PDFs of $g_\ell$'s, $\ell=0,1,\ldots,L$, are given by
\begin{equation}\label{Eq:PDF_link}
f_{g_{\ell}}\left(x\right) = \mathcal{P}_\ell\delta\left(x\right)+\frac{\left(1-\mathcal{P}_\ell\right)C_{\ell}^{m_{\ell}}}{\Gamma\left(m_\ell\right)}x^{m_{\ell}-1}\exp\left(-C_{\ell}x\right)
\end{equation}
where $\mathcal{P}_0$ is the probability that $D$ belongs to $\mathbb{B}$. Clearly, the direct $S \longrightarrow D$ path is not linked via a relay, i.e., $\mathcal{P}_{0}=0$, yielding $f_{g_{0}}\left(x\right)=f_{\gamma_{0}}\left(x\right)$. Integrating \eqref{Eq:PDF_link} and using \eqref{Eq:Gamma_CDF}, yields the following expression for the CDF of the $\ell$-th cascaded path:
\begin{equation}\label{Eq:CDF_link}
F_{g_{\ell}}\left(x\right) = \mathcal{P}_\ell u\left(x\right)+\left(1-\mathcal{P}_\ell\right)\frac{\mathsf{\Gamma}\left(m_\ell,C_{\ell}x\right)}{\Gamma\left(m_\ell\right)}.
\end{equation}
Note again that for the direct $S \longrightarrow D$ path yields $F_{g_{0}}\left(x\right)=F_{\gamma_{0}}\left(x\right)$.

\subsection{Repetitive Transmission}\label{Sec:Repetitive}
The incoming signals at $D$ from $S$ and $R_k\in\mathbb{B}$ $\forall\,k$ may be combined using a time-diversity version of MRD \cite{J:Laneman_2003} and SD \cite{J:Hu_DF_2007}. In particular, $D$ combines $S$'s signal received at time slot one with the $S$'s replicas received from all $R_k\in\mathbb{B}$ $\forall\,k$ at the $L$ subsequent slots using either the rules of MRD or SD.

\subsubsection{Repetitive with MRD}\label{Sec:MRD_Rep}
With MRD the instantaneous SNR at $D$'s output is expressed as
\begin{equation}\label{Eq:g_end_MRD}
g_{\text{end}} = g_0+\sum_{i=1}^{\left|\mathbb{B}\right|}g_i.
\end{equation}
Since $g_\ell$'s, $\ell=0,1,\ldots,L$, are independent, the moment generating function (MGF) of $g_{\text{end}}$ can be easily obtained as the product of the MGFs
of $g_\ell$'s. As shown in \cite{J:Datsikas_2008} for INID Nakagami-$m$ fading, using \eqref{Eq:PDF_link} and the definition of the MGF of $g_k$, $k=1,2,\ldots,L$,
\begin{equation}\label{Eq:MGF_Definition}
M_{g_k}\left(s\right) = \int_0^\infty\exp\left(-sx\right)f_{g_k}\left(x\right)dx,
\end{equation}
yields $\mathcal{M}_{g_{k}}\left(s\right)=\mathcal{P}_k+\left(1-\mathcal{P}_k\right)C_k^{m_k}\left(s+C_k\right)^{-m_k}$. Similarly, using \eqref{Eq:Gamma_PDF} the MGF of $g_0$ is easily obtained as $\mathcal{M}_{g_{0}}\left(s\right)=C_0^{m_0}\left(s+C_0\right)^{-m_0}$. Hence, the following closed-form expression for the MGF of $g_{\text{end}}$ in INID Nakagami-$m$ is deduced
\begin{equation}\label{Eq:MGF_end_Def_Rep}
\mathcal{M}_{g_{\text{end}}}\left(s\right) = C_0^{m_0}\left(s+C_0\right)^{-m_0}\prod_{k=1}^L\left[\mathcal{P}_k+
\left(1-\mathcal{P}_k\right)C_k^{m_k}\left(s+C_k\right)^{-m_k}\right].
\end{equation}
Using the MGF-based approach \cite{B:Sim_Alou_Book}, the CDF of $g_{\text{end}}$ can be obtained as
\begin{equation}\label{Eq:Inverse_Laplaca}
F_{g_{\text{end}}}\left(x\right)=\mathbb{L}^{-1}\left\{\frac{\mathcal{M}_{g_{\text{end}}}\left(s\right)}{s};x\right\}.
\end{equation}
Substituting \eqref{Eq:MGF_end_Def_Rep} in \eqref{Eq:Inverse_Laplaca} and similar to the analysis presented in \cite{J:Datsikas_2008}, a closed-form expression for $F_{g_{\text{end}}}\left(x\right)$ of repetitive transmission with MRD over INID Nakagami-$m$ with integer $m_\ell$'s and distinct $C_\ell$'s is given by
\begin{equation}\label{Eq:CDF_gEND_MRD_Distinct_C}
\begin{split}
F_{g_{\text{end}}}\left(x\right) =& \left(\prod_{\ell=0}^L\mathcal{P}_\ell\right)\left\{1
+\sum_{\left\{\lambda_k\right\}_{k=0}^{L}}\left(\prod_{n=0}^k
\frac{1-\mathcal{P}_{\lambda_n}}{\mathcal{P}_{\lambda_n}}C_{\lambda_n}^{m_{\lambda_n}}\right)\right.
\\&\left.\times\sum_{p=0}^k\sum_{q=1}^{m_{\lambda_p}}
\frac{\psi_p}{C_{\lambda_p}^{q}}\left[1-\exp\left(-C_{\lambda_p}x\right)\sum_{\ell=0}^{q-1}\frac{\left(C_{\lambda_p}x\right)^\ell}{\ell!}\right]\right\}
\end{split}
\end{equation}
The symbol $\sum_{\left\{\alpha_i\right\}_{i=\kappa}^{I}}$ is used for short-hand representation of multiple summations $\sum_{i=\kappa}^{I}$ $\sum_{\alpha_\kappa=\kappa}^{I-i+\kappa}$ $\sum_{\alpha_{\kappa+1}=\alpha_\kappa+1}^{I-i+\kappa+1}$ $\cdots$ $\sum_{\alpha_i=\alpha_{i-1}+1}^{I}$ and $\psi_{p}=\Psi_p\left(s\right)^{(m_{\lambda_p}-q)}|_{s=-C_{\lambda_p}}/(m_{\lambda_p}-q)!$ with $\Psi_p\left(s\right)=(s+C_{\lambda_p})^{m_{\lambda_p}}\prod_{n=0}^k(s+C_{\lambda_n})^{-m_{\lambda_n}}$. For $C_\ell=C$ $\forall\,\ell$ with arbitrary values for $m_\ell$'s and following a similar analysis as for the derivation of \eqref{Eq:CDF_gEND_MRD_Distinct_C}, a closed-form expression for $F_{g_{\text{end}}}\left(x\right)$ can be obtained as
\begin{equation}\label{Eq:CDF_gEND_MRD_Same_C}
F_{g_{\text{end}}}\left(x\right) = \left(\prod_{\ell=0}^L\mathcal{P}_\ell\right)\left\{1
+\sum_{\left\{\lambda_k\right\}_{k=0}^{L}}\left(\prod_{n=0}^k
\frac{1-\mathcal{P}_{\lambda_n}}{\mathcal{P}_{\lambda_n}}\right)\mathsf{\Gamma}\left(\sum_{n=0}^km_{\lambda_n},Cx\right)/\Gamma\left(\sum_{n=0}^km_{\lambda_n}\right)\right\}.
\end{equation}

\subsubsection{Repetitive with SD}\label{Sec:SD_Rep}
Alternatively to MRD, $D$ may use a time-diversity version of SD to combine the signals from $S$ and $R_k\in\mathbb{B}$ $\forall\,k$ \cite{J:Hu_DF_2007}. With this diversity technique the instantaneous SNR at $D$'s output is given by
\begin{equation}\label{Eq:g_end_SD}
g_{\text{end}} = \max_{R_k\in\mathbb{C}}g_k
\end{equation}
where $\mathbb{C} = \{S\} \cup \mathbb{B}$. The $g_\ell$'s, $\ell=0,1,\ldots,L$, are assumed independent, therefore $F_{g_{\text{end}}}\left(x\right) $ of \eqref{Eq:g_end_SD} can be easily obtained as the product of the CDFs of $g_\ell$'s. Substituting \eqref{Eq:Gamma_CDF} and \eqref{Eq:CDF_link} for $g_0$ and $g_k$ $\forall\,k=1,2,\ldots,L$, respectively, a closed-form expression for $F_{g_{\text{end}}}\left(x\right) $ of repetitive transmission with SD over INID Nakagami-$m$ fading can be derived as
\begin{equation}\label{Eq:CDF_gEND_SD_Distinct_C}
F_{g_{\text{end}}}\left(x\right) = F_{\gamma_0}\left(x\right)F_{g_{\text{best}}}\left(x\right)
\end{equation}
where $F_{g_{\text{best}}}\left(x\right)$ is the CDF of the instantaneous SNR of the $R_{\rm best} \rightarrow D$ channel, i.e., of
\begin{equation}\label{Eq:g_best}
g_{\text{best}} = \max_{R_k\in\mathbb{B}}g_k,
\end{equation}
which is easily obtained using \eqref{Eq:CDF_link} for INID Nakagami-$m$ fading as
\begin{equation}\label{Eq:CDF_gbest_SD_Distinct_C}
F_{g_{\text{best}}}\left(x\right) = \prod_{k=1}^L\left[\mathcal{P}_ku\left(x\right)+\left(1-\mathcal{P}_k\right)
\frac{\mathsf{\Gamma}\left(m_k,C_{k}x\right)}{\Gamma\left(m_k\right)}\right].
\end{equation}

Differentiating \eqref{Eq:CDF_gEND_SD_Distinct_C}, the PDF of $g_{\text{end}}$ is given by
\begin{equation}\label{Eq:PDF_gEND_SD_Distinct_C}
f_{g_{\text{end}}}\left(x\right) = f_{\gamma_0}\left(x\right)F_{g_{\text{best}}}\left(x\right)+F_{\gamma_0}\left(x\right)f_{g_{\text{best}}}\left(x\right)
\end{equation}
where $f_{g_{\text{best}}}\left(x\right)$ is the PDF of $g_{\text{best}}$. To obtain an expression for $f_{g_{\text{best}}}\left(x\right)$, we first use \cite[eq. (8.352/1)]{B:Gra_Ryz_Book} to obtain \eqref{Eq:CDF_gbest_SD_Distinct_C} for integer $m_\ell$'s yielding
\begin{equation}\label{Eq:CDF_gbest_SD_Distinct_C_integerm}
F_{g_{\text{best}}}\left(x\right) =\prod_{k=1}^L\left\{\mathcal{P}_ku\left(x\right)+\left(1-\mathcal{P}_k\right)
\left[1-\exp\left(-C_kx\right)\sum_{i=0}^{m_k-1}\frac{\left(C_kx\right)^i}{i!}\right]\right\}.
\end{equation}
Then, differentiating \eqref{Eq:CDF_gbest_SD_Distinct_C_integerm} and using the formula
\begin{equation}\label{Eq:formula}
\prod_{i=\kappa}^I\left(\chi_i+\psi_i\right) = \prod_{i=\kappa}^I\chi_i+\sum_{\left\{\alpha_i\right\}_{i=\kappa}^{I}}
\prod_{s=\kappa}^i\psi_{\alpha_s}\prod_{\substack{t=\kappa \\ t\neq \left\{\alpha_u\right\}_{u=\kappa}^{i}}}^I\chi_t,
\end{equation}
where symbol $\sum_{\left\{\alpha_i\right\}_{i=\kappa}^{I}}$ is used for short-hand representation of multiple summations $\sum_{i=\kappa}^{I}$ $\sum_{\alpha_\kappa=\kappa}^{I-i+\kappa}\sum_{\alpha_{\kappa+1}=\alpha_\kappa+1}^{I-i+\kappa+1}\cdots\sum_{\alpha_i=\alpha_{i-1}+1}^{I}$, we obtain after some algebraic manipulations the following closed-form expression for $f_{g_{\text{best}}}\left(x\right)$ in INID Nakagami-$m$ fading with integer $m_\ell$'s:
\begin{equation}\label{Eq:PDF_best_Distinct}
\begin{split}
&f_{g_{\text{best}}}\left(x\right) = \left(\prod_{k=1}^L\mathcal{P}_k\right)
\left\{\delta\left(x\right)+\sum_{\left\{\lambda_k\right\}_{k=1}^{L}}\sum_{\left\{\mu_j\right\}_{j=1}^{k}}
\left(-1\right)^j\left(\prod_{n=1}^k
\frac{1-\mathcal{P}_{\lambda_n}}{\mathcal{P}_{\lambda_n}}\right)\right.
\\&\left.\times
\left[-\left(\sum_{p=1}^jC_{\lambda_{\mu_p}}\right)\exp\left(-x\sum_{p=1}^jC_{\lambda_{\mu_p}}\right)+\sum_{\left\{\nu_i\right\}_{i=1}^{j}}
\left(\sum_{s=1}^pi_{\nu_s}\right)\left(\prod_{s=1}^p
\sum_{i_{\nu_s}=1}^{m_{\lambda_{\mu_{\nu_s}}}-1}\frac{C_{\lambda_{\mu_{\nu_s}}}^{i_{\nu_s}}}{i_{\nu_s}!}\right)\right.\right.
\\&\left.\left.\times x^{\sum_{s=1}^pi_{\nu_s}-1}\exp\left(-x\xi_{p,j}\right)\left(1-\frac{\xi_{p,j}x}{\sum_{s=1}^pi_{\nu_s}}\right)\right]\right\}.
\end{split}
\end{equation}
In \eqref{Eq:PDF_best_Distinct}, parameters $\xi_{p,j}$'s, with $p$ and $j$ being positive integers, are given by
\begin{equation}\label{Eq:xi}
\xi_{p,j}= \sum_{s=1}^pC_{\lambda_{\mu_{\nu_s}}}+\sum_{\substack{t=1 \\ t\neq \left\{\nu_u\right\}_{u=1}^{p}}}^jC_{\lambda_{\mu_t}}.
\end{equation}
Substituting \eqref{Eq:Gamma_PDF} and \eqref{Eq:Gamma_CDF} for integer $m_\ell$'s, i.e., after using \cite[eq. (8.352/1)]{B:Gra_Ryz_Book} for expressing $\mathsf{\Gamma}\left(\cdot,\cdot\right)$'s, as well as \eqref{Eq:CDF_gbest_SD_Distinct_C_integerm} and \eqref{Eq:PDF_best_Distinct} to \eqref{Eq:PDF_gEND_SD_Distinct_C}, a closed-form expression for $f_{g_{\text{end}}}\left(x\right)$ over INID Nakagami-$m$ fading with integer values of $m_\ell$'s and distinct $C_\ell$'s can be obtained after some algebraic manipulations as
\begin{equation}\label{Eq:PDF_gEND_SD_Distinct_C_FINAL}
\begin{split}
&f_{g_{\text{end}}}\left(x\right) = \left(\prod_{k=1}^L\mathcal{P}_k\right)\left\{\frac{C_{0}^{m_{0}}}{\Gamma\left(m_0\right)}x^{m_{0}-1}\exp\left(-C_{0}x\right)
\left\{u\left(x\right)+\sum_{\left\{\lambda_k\right\}_{k=1}^{L}}\sum_{\left\{\mu_j\right\}_{j=1}^{k}}
\left(-1\right)^j\right.\right.
\\&\left.\left.\times\left(\prod_{n=1}^k
\frac{1-\mathcal{P}_{\lambda_n}}{\mathcal{P}_{\lambda_n}}\right)
\left[\exp\left(-x\sum_{p=1}^jC_{\lambda_{\mu_p}}\right)-1+\sum_{\left\{\nu_i\right\}_{i=1}^{j}}
\left(\prod_{s=1}^p
\sum_{i_{\nu_s}=1}^{m_{\lambda_{\mu_{\nu_s}}}-1}\frac{C_{\lambda_{\mu_{\nu_s}}}^{i_{\nu_s}}}{i_{\nu_s}!}\right)\right.\right.\right.
\\&\left.\left.\left.\times x^{\sum_{s=1}^pi_{\nu_s}}\exp\left(-x\xi_{p,j}\right)\right]\right\}+
\left[1-\exp\left(-C_0x\right)\sum_{i=0}^{m_0-1}\frac{\left(C_0x\right)^i}{i!}\right]\right.
\\&\left.\times
\left\{\delta\left(x\right)+\sum_{\left\{\lambda_k\right\}_{k=1}^{L}}\sum_{\left\{\mu_j\right\}_{j=1}^{k}}
\left(-1\right)^j\left(\prod_{n=1}^k
\frac{1-\mathcal{P}_{\lambda_n}}{\mathcal{P}_{\lambda_n}}\right)\right.\right.
\\&\left.\left.\times
\left[-\left(\sum_{p=1}^jC_{\lambda_{\mu_p}}\right)\exp\left(-x\sum_{p=1}^jC_{\lambda_{\mu_p}}\right)+\sum_{\left\{\nu_i\right\}_{i=1}^{j}}
\left(\sum_{s=1}^pi_{\nu_s}\right)\left(\prod_{s=1}^p
\sum_{i_{\nu_s}=1}^{m_{\lambda_{\mu_{\nu_s}}}-1}\frac{C_{\lambda_{\mu_{\nu_s}}}^{i_{\nu_s}}}{i_{\nu_s}!}\right)\right.\right.\right.
\\&\left.\left.\left.\times x^{\sum_{s=1}^pi_{\nu_s}-1}\exp\left(-x\xi_{p,j}\right)\left(1-\frac{\xi_{p,j}x}{\sum_{s=1}^pi_{\nu_s}}\right)\right]\right\}\right\}.
\end{split}
\end{equation}

To obtain the MGF of $g_{\text{end}}$, we substitute the $f_{g_{\text{end}}}\left(x\right)$ expression given by \eqref{Eq:PDF_gEND_SD_Distinct_C_FINAL} to the definition of the MGF given by \eqref{Eq:MGF_Definition}, i.e., after replacing $g_k$ with $g_{\text{end}}$ in \eqref{Eq:MGF_Definition}, and use\cite[eq. (3.381/4)]{B:Gra_Ryz_Book} to solve the resulting integrals. In particular, by first deriving using \eqref{Eq:PDF_best_Distinct} the following closed-form expression for the MGF of $g_{\text{best}}$ in INID Nakagami-$m$ fading with integer values of $m_\ell$'s and distinct $C_\ell$'s:
\begin{equation}\label{Eq:MGF_gbest}
\begin{split}
&M_{g_{\text{best}}}\left(s\right) =
\left(\prod_{k=1}^L\mathcal{P}_k\right)
\left\{1+\sum_{\left\{\lambda_k\right\}_{k=1}^{L}}\sum_{\left\{\mu_j\right\}_{j=1}^{k}}\left(-1\right)^j\left(\prod_{n=1}^k
\frac{1-\mathcal{P}_{\lambda_n}}{\mathcal{P}_{\lambda_n}}\right)\right.
\\&\times\left.\left\{-\left(\sum_{p=1}^jC_{\lambda_{\mu_p}}\right)\left(s+\sum_{p=1}^jC_{\lambda_{\mu_p}}\right)^{-1}
+\sum_{\left\{\nu_i\right\}_{i=1}^{j}}
\left(\sum_{s=1}^pi_{\nu_s}\right)!\left(\prod_{s=1}^p
\sum_{i_{\nu_s}=1}^{m_{\lambda_{\mu_{\nu_s}}}-1}\frac{C_{\lambda_{\mu_{\nu_s}}}^{i_{\nu_s}}}{i_{\nu_s}!}\right)\right.\right.
\\&\left.\left.\times\left(s+\xi_{p,j}\right)^{-\sum_{s=1}^pi_{\nu_s}}\left[1-\xi_{p,j}\left(s+\xi_{p,j}\right)^{-1}\right]\right\}\right\},
\end{split}
\end{equation}
a closed-form expression for the MGF of $g_{\text{end}}$ of repetitive transmission with SD over INID Nakagami-$m$ fading channels with integer values of $m_\ell$'s and distinct $C_\ell$'s is given by
\begin{equation}\label{Eq:MGF_end_Final_repSD}
\begin{split}
&\mathcal{M}_{g_{\text{end}}}\left(s\right) = \left(\prod_{k=1}^L\mathcal{P}_k\right)\left\{C_{0}^{m_{0}}
\left\{\left(s+C_0\right)^{-m_0}+\sum_{\left\{\lambda_k\right\}_{k=1}^{L}}\sum_{\left\{\mu_j\right\}_{j=1}^{k}}
\left(-1\right)^j\left(\prod_{n=1}^k
\frac{1-\mathcal{P}_{\lambda_n}}{\mathcal{P}_{\lambda_n}}\right)\right.\right.
\\&\left.\left.\times
\left[\left(s+C_0+\sum_{p=1}^jC_{\lambda_{\mu_p}}\right)^{-m_0}-\left(s+C_0\right)^{-m_0}+\sum_{\left\{\nu_i\right\}_{i=1}^{j}}
\left(\prod_{s=1}^p
\sum_{i_{\nu_s}=1}^{m_{\lambda_{\mu_{\nu_s}}}-1}\frac{C_{\lambda_{\mu_{\nu_s}}}^{i_{\nu_s}}}{i_{\nu_s}!}\right)\right.\right.\right.
\\&\left.\left.\left.\times\frac{\left(m_0+\sum_{s=1}^pi_{\nu_s}-1\right)!}{\left(m_0-1\right)!}
\left(s+C_0+\xi_{p,j}\right)^{-\left(m_0+\sum_{s=1}^pi_{\nu_s}\right)}\right]\right\}+\left\{\sum_{\left\{\lambda_k\right\}_{k=1}^{L}}\sum_{\left\{\mu_j\right\}_{j=1}^{k}}
\left(-1\right)^j\right.\right.
\\&\left.\left.\left(\prod_{n=1}^k
\frac{1-\mathcal{P}_{\lambda_n}}{\mathcal{P}_{\lambda_n}}\right)\left\{\left(\sum_{p=1}^jC_{\lambda_{\mu_p}}\right)\sum_{q=0}^{m_0-1}C_0^q
\left(s+C_0+\sum_{p=1}^jC_{\lambda_{\mu_p}}\right)^{-(q+1)}\right.\right.\right.
\\&\left.\left.\left.+\sum_{\left\{\nu_i\right\}_{i=1}^{j}}
\left(\sum_{s=1}^pi_{\nu_s}\right)\left(\prod_{s=1}^p
\sum_{i_{\nu_s}=1}^{m_{\lambda_{\mu_{\nu_s}}}-1}\frac{C_{\lambda_{\mu_{\nu_s}}}^{i_{\nu_s}}}{i_{\nu_s}!}\right)
\sum_{q=0}^{m_0-1}\frac{C_0^q}{q!}\left(s+C_0+\xi_{p,j}\right)^{-\left(q+\sum_{s=1}^pi_{\nu_s}+1\right)}\right.\right.\right.
\\&\left.\left.\left.\times \left(q+\sum_{s=1}^pi_{\nu_s}-1\right)!
\left[q\xi_{p,j}-\left(s+C_0\right)\sum_{s=1}^pi_{\nu_s}\right]\right\}\right\}\right\}
+M_{g_{\text{best}}}\left(s\right).
\end{split}
\end{equation}
For equal $C_\ell$'s, i.e., IID Nakagami-$m$ fading with $m_\ell=m$ and $\overline{\gamma}_\ell=\overline{\gamma}$ $\forall\,\ell$, following a similar procedure as for the derivation of \eqref{Eq:MGF_end_Final_repSD} and using the binomial and multinomial theorems \cite[eq. (1.111)]{B:Gra_Ryz_Book}, we first obtain the following closed-form expression for $M_{g_{\text{best}}}\left(s\right)$ for integer $m$
\begin{equation}\label{Eq:MGF_gbest_EqualC}
\begin{split}
&M_{g_{\text{best}}}\left(s\right) =
L\mathcal{P}^L
\left\{1/L+\left(1-\mathcal{P}\right)C^m\left(s+C\right)^{-m}+\sum_{k=1}^{L-1}\binom{L-1}{k}\frac{\left(1-\mathcal{P}\right)^k}{\mathcal{P}^{k+1}}\right.
\\&\left.\times\left\{\mathcal{P}+\left(1-\mathcal{P}\right)C^m\left(s+C\right)^{-m}+\sum_{j=1}^k\binom{k}{j}\left(-1\right)^j\right.\right.
\\&\left.\left.\times\left[\mathcal{P}+\sum_{n_1+n_2+\cdots+n_m=j}^j\frac{\left(1-\mathcal{P}\right)C^{m+\sigma}}{\prod_{j=2}^{m-1}\left(j!\right)^{n_{j+1}}}
\frac{\prod_{i=1}^mn_i^{-1}}{\left(m-1\right)!}\left[s+(j+1)C\right]^{-\left(m+
\sigma\right)}\right]\right\}\right\}
\end{split}
\end{equation}
where $C=m/\overline{\gamma}$, $\mathcal{P}=\mathcal{P}_k$ $\forall\,k$, $\sigma=\sum_{i=2}^m(i-1)n_i$ and symbol $\sum_{n_1+n_2+\cdots+n_m=j}^j$ is used for short-hand representation of multiple summations $\sum_{n_1=0}^j\sum_{n_2=0}^j\cdots\sum_{n_m=0}^j\delta(\sum_{i=1}^mn_i,j)$. Using \eqref{Eq:PDF_gEND_SD_Distinct_C} for IID Nakagami-$m$ fading, \eqref{Eq:MGF_gbest_EqualC} and after some algebraic manipulations, a closed-form expression for $M_{g_{\text{end}}}\left(s\right)$ of repetitive transmission with SD in IID Nakagami-$m$ fading channels with integer $m$ is obtained as
\begin{equation}\label{Eq:MGF_gend_FINAL_EqualC}
\begin{split}
&M_{g_{\text{end}}}\left(s\right) =
L\mathcal{P}^L\left\{C^m\left\{\left[1/L+\left(1-\mathcal{P}\right)\right]\left(s+C\right)^{-m}-(1-\mathcal{P})B_{0,1}\left(s\right)+
\sum_{k=1}^{L-1}\binom{L-1}{k}\right.\right.
\\&\left.\left.\times\frac{\left(1-\mathcal{P}\right)^k}{\mathcal{P}^{k+1}}
\left\{\left[\mathcal{P}+\left(1-\mathcal{P}\right)\right]\left(s+C\right)^{-m}
-(1-\mathcal{P})B_{0,1}\left(s\right)+\sum_{j=1}^k\binom{k}{j}\left(-1\right)^j\left[\mathcal{P}\right.\right.\right.\right.
\\&\left.\left.\left.\left.\times\left(s+C\right)^{-m}+\sum_{n_1+n_2+\cdots+n_m=j}^j
\frac{\left(1-\mathcal{P}\right)}{\prod_{j=2}^{m-1}\left(j!\right)^{n_{j+1}}}
\frac{\prod_{i=1}^m n_i^{-1}}{\left(m-1\right)!}\left\{\left(s+C\right)^{-m}-B_{\sigma,j+1}\left(s\right)\right\}\right]\right\}\right\}\right.
\\&\left.-C^m\left\{\left(1-\mathcal{P}\right)B_{0,1}\left(s\right)+\sum_{k=1}^{L-1}\binom{L-1}{k}
\frac{\left(1-\mathcal{P}\right)^k}{\mathcal{P}^{k+1}}
\left\{\left(1-\mathcal{P}\right)B_{0,1}\left(s\right)+\sum_{j=1}^k\binom{k}{j}\left(-1\right)^j\right.\right.\right.
\\&\left.\left.\left.\times\sum_{n_1+n_2+\cdots+n_m=j}^j\frac{\left(1-\mathcal{P}\right)C^\sigma}{\prod_{j=2}^{m-1}
\left(j!\right)^{n_{j+1}}}
\frac{\prod_{i=1}^m n_i^{-1}}{\left(m-1\right)!}\sum_{q=0}^{m-1}\frac{C^{q}}{q!(m+\sigma-1)!}(m+q+\sigma-1)!\right.\right.\right.
\\&\left.\left.\left.\times[s+(\lambda+1)C]^{-(m+q)}\right\}\right\}\right\}+M_{g_{\text{best}}}\left(s\right)
\end{split}
\end{equation}
where function $B_{\kappa,\lambda}\left(s\right)$, with $\kappa$ and $\lambda$ being positive integers, is given by
\begin{equation}\label{Eq:Function_B}
B_{\kappa,\lambda}\left(s\right) = \sum_{q=0}^{m+\kappa-1}\frac{(\lambda C)^{q}}{q!(m-1)!}(m+q-1)![s+(\lambda+1)C]^{-(m+q)}.
\end{equation}

\subsection{RS-based Transmission}\label{Sec:RS-based}
When RS-based transmission is utilized, RS \cite{J:Bletsas_2006} is first performed to obtain $R_{\rm best}$. Relay node $R_{\rm best}$ is the one experiencing the most favorable $R_{k} \rightarrow D$ $\forall\,k$ channel conditions, i.e., its instantaneous SNR is given by \eqref{Eq:g_best}. Using expressions derived in Section~\ref{Sec:Repetitive}, we next obtain closed-form expressions for the statistics of a pure RS scheme \cite{J:Beres_2008, J:Geo_Agi_RS_2010} that combines at $D$ the signals from $S$ and $R_{\rm best}$ using a time-diversity version of MRD as well as of a rate-selective one \cite{J:Geo_Agi_RS_2010} that utilizes pure RS only if it is beneficial in terms of achievable rate over the direct transmission.

\subsubsection{Pure RS}\label{Sec:FSDF}
With pure RS $D$ utilizes MRD to combine the signals from $S$ and $R_{\rm best}$ \cite{J:Beres_2008}, therefore the instantaneous SNR at $D$'s output is given using \eqref{Eq:g_best} by
\begin{equation}\label{Eq:g_end_PureRS}
g_{\text{end}} = g_0+g_{\text{best}}.
\end{equation}
Similar to the derivation of \eqref{Eq:MGF_end_Def_Rep}, the MGF of $g_{\text{end}}$ of pure RS can be obtained as the following product:
\begin{equation}\label{Eq:MGF_g_end_PureRS}
\mathcal{M}_{g_{\text{end}}}\left(s\right) = C_0^{m_0}\left(s+C_0\right)^{-m_0}\mathcal{M}_{g_{\text{best}}}\left(s\right)
\end{equation}
where $\mathcal{M}_{g_{\text{best}}}\left(s\right)$ is given by \eqref{Eq:MGF_gbest} for INID Nakagami-$m$ with integer $m_\ell$'s and distinct $C_\ell$'s, whereas by \eqref{Eq:MGF_gbest_EqualC} for IID Nakagami-$m$ with integer $m_\ell=m$ $\forall\,\ell$ and $C_\ell=C$. Therefore substituting \eqref{Eq:MGF_gbest} to \eqref{Eq:MGF_g_end_PureRS}, a closed-form expression for $\mathcal{M}_{g_{\text{end}}}\left(s\right)$ of pure RS for INID Nakagami-$m$ fading with integer $m_\ell$'s and distinct $C_\ell$'s is given by \cite[eq. (7)]{J:Geo_Agi_RS_2010}
\begin{equation}\label{Eq:MGF_end_Final}
\begin{split}
&\mathcal{M}_{g_{\text{end}}}\left(s\right) =
C_0^{m_0}\left(s+C_0\right)^{-m_0}\left(\prod_{k=1}^L\mathcal{P}_k\right)
\left\{1+\sum_{\left\{\lambda_k\right\}_{k=1}^{L}}\sum_{\left\{\mu_j\right\}_{j=1}^{k}}\left(-1\right)^j\left(\prod_{n=1}^k
\frac{1-\mathcal{P}_{\lambda_n}}{\mathcal{P}_{\lambda_n}}\right)\right.
\\&\times\left.\left\{-\left(\sum_{p=1}^jC_{\lambda_{\mu_p}}\right)\left(s+\sum_{p=1}^jC_{\lambda_{\mu_p}}\right)^{-1}
+\sum_{\left\{\nu_i\right\}_{i=1}^{j}}
\left(\sum_{s=1}^pi_{\nu_s}\right)!\left(\prod_{s=1}^p
\sum_{i_{\nu_s}=1}^{m_{\lambda_{\mu_{\nu_s}}}-1}\frac{C_{\lambda_{\mu_{\nu_s}}}^{i_{\nu_s}}}{i_{\nu_s}!}\right)\right.\right.
\\&\left.\left.\times\left(s+\xi_{p,j}\right)^{-\sum_{s=1}^pi_{\nu_s}}\left[1-\xi_{p,j}\left(s+\xi_{p,j}\right)^{-1}\right]\right\}\right\}.
\end{split}
\end{equation}
For IID Nakagami-$m$ fading channels with integer $m$, substituting \eqref{Eq:MGF_gbest_EqualC} to \eqref{Eq:MGF_g_end_PureRS} a closed-form expression for $\mathcal{M}_{g_{\text{end}}}\left(s\right)$ of pure RS is given by \cite[eq. (10)]{J:Geo_Agi_RS_2010}
\begin{equation}\label{Eq:MGF_end_Final_Equalm}
\begin{split}
&\mathcal{M}_{g_{\text{end}}}\left(s\right) = L\mathcal{P}^LC^m\left(s+C\right)^{-m}
\left\{\frac{1}{L}+\left(1-\mathcal{P}\right)C^m\left(s+C\right)^{-m}+\sum_{k=1}^{L-1}\binom{L-1}{k}\frac{\left(1-\mathcal{P}\right)^k}{\mathcal{P}^{k+1}}\right.
\\&\left.\times\left\{\mathcal{P}+\left(1-\mathcal{P}\right)C^m\left(s+C\right)^{-m}+\sum_{j=1}^k\binom{k}{j}
\left(-1\right)^j\left[\mathcal{P}+\sum_{n_1+n_2+\cdots+n_m=j}^j\frac{\left(1-\mathcal{P}\right)C^{m+\sigma}}{\prod_{j=2}^{m-1}\left(j!\right)^{n_{j+1}}}\right.\right.\right.
\\&\left.\left.\left.\times\frac{\prod_{i=1}^mn_i^{-1}}{\left(m-1\right)!}\left[s+(j+1)C\right]^{-\left(m+
\sigma\right)}\right]\right\}\right\}.
\end{split}
\end{equation}

To obtain $F_{g_{\text{end}}}\left(x\right)$ of pure RS for INID Nakagami-$m$ fading with integer $m_\ell$'s and distinct $C_\ell$'s, we substitute \eqref{Eq:MGF_end_Final} to \eqref{Eq:Inverse_Laplaca} and after some algebraic manipulations yields the following closed-form expression \cite[eq. (10)]{J:Geo_Agi_RS_2010}
\begin{equation}\label{Eq:OP_FSDF_Difm}
\begin{split}
&F_{g_{\text{end}}}\left(x\right) = \left(\prod_{k=1}^L\mathcal{P}_k\right)\left\{\mathcal{Z}_{m_0}\left(C_0,x\right)
+\sum_{\left\{\lambda_k\right\}_{k=1}^{L}}\sum_{\left\{\mu_j\right\}_{j=1}^{k}}\left(-1\right)^j\left(\prod_{n=1}^k
\frac{1-\mathcal{P}_{\lambda_n}}{\mathcal{P}_{\lambda_n}}\right)\right.
\\&\left.\left.\left\{-C_0^{m_0}\left\{\mathcal{X}_{2,1}^{(1)}\mathcal{Z}_{0}\left[\chi_2^{(1)},x\right]
+\chi_2^{(1)}\mathcal{Y}_1^{(1)}\left(x\right)\right\}+\sum_{\left\{\nu_i\right\}_{i=1}^{j}}\left(\prod_{s=1}^p
\sum_{i_{\nu_s}=1}^{m_{\lambda_{\mu_{\nu_s}}}-1}\frac{C_{\lambda_{\mu_{\nu_s}}}^{i_{\nu_s}}}{i_{\nu_s}!}\right)\right.\right.\right.
\\&\left.\left.\times C_0^{m_0}\left(\sum_{s=1}^pi_{\nu_s}\right)!\left[\mathcal{Y}_2^{(2)}\left(x\right)-\xi_{p,j}\mathcal{Y}_2^{(3)}\left(x\right)\right]\right\}\right\}
\end{split}
\end{equation}
where $\mathcal{Z}_t\left(c,x\right) = 1-\exp\left(-cx\right)\sum_{i=1}^{t-1}\left(cx\right)^i/i!$, with $c$ being positive real, and
\begin{equation}\label{Eq:Y}
\mathcal{Y}_\kappa^{(u)}\left(x\right) = \sum_{r=1}^\kappa\sum_{t=1}^{b_r^{(u)}}\mathcal{X}_{r,t}^{(u)}\left[\chi_r^{(u)}\right]^{-t}\mathcal{Z}_t\left[\chi_r^{(u)},x\right]
\end{equation}
for $u=1,2$ and $3$. In the above two equations $\forall\,u$, $\mathcal{X}_{r,t}^{(u)}=X_r^{(u)}\left(s\right)^{(b_r^{(u)}-t)}|_{s=-\chi_r^{(u)}}/(b_r^{(u)}-t)!$ and  $X_r^{(u)}\left(s\right)=(s+\chi_r^{(u)})^{b_r^{(u)}}\prod_{j=1}^2(s+\chi_j^{(u)})^{b_j^{(u)}}$. Moreover, $\chi_1^{(u)}=C_0$ $\forall\,u$, $\chi_2^{(1)}=\sum_{p=1}^jC_{\lambda_{\mu_p}}$ and $\chi_2^{(2)}=\chi_2^{(3)}=\xi_{p,j}$ as well as $b_1^{(u)}=m_0$, $\forall\,u$, $b_2^{(1)}=1$, $b_2^{(2)}=\sum_{s=1}^pi_{\nu_s}$ and $b_2^{(3)}=b_2^{(2)}+1$. For $C_\ell=C$ $\forall\,\ell$ and integer $m$ and similar to the derivation of \eqref{Eq:OP_FSDF_Difm}, substituting \eqref{Eq:MGF_end_Final_Equalm} to \eqref{Eq:Inverse_Laplaca}, a closed-form expression for $F_{g_{\text{end}}}\left(x\right)$ of pure RS is given by \cite[eq. (11)]{J:Geo_Agi_RS_2010}
\begin{equation}\label{Eq:OP_FSDF_Equalm}
\begin{split}
&F_{g_{\text{end}}}\left(x\right) = L\mathcal{P}^L
\left\{\frac{\mathcal{Z}_{m}\left(C,x\right)}{L}+\left(1-\mathcal{P}\right)\mathcal{Z}_{2m}\left(C,x\right)
+\sum_{k=1}^{L-1}\binom{L-1}{k}\frac{\left(1-\mathcal{P}\right)^k}{\mathcal{P}^{k+1}}
\right.
\\&\left.\left\{\mathcal{P}\mathcal{Z}_{m}\left(C,x\right)+\left(1-\mathcal{P}\right)\mathcal{Z}_{2m}\left(C,x\right)+\sum_{j=1}^k\binom{k}{j}
\left(-1\right)^j\left[\mathcal{P}\mathcal{Z}_{m}\left(C,x\right)\right.\right.\right.
\\&\left.\left.\left.+\sum_{n_1+n_2+\cdots+n_m=j}^j\left(1-\mathcal{P}\right)
\frac{\prod_{i=1}^mn_i^{-1}C^{2m+\sigma}}{\left(m-1\right)!\prod_{j=2}^{m-1}\left(j!\right)^{n_{j+1}}}\mathcal{Y}_2^{(4)}
\left(x\right)\right]\right\}\right\}
\end{split}
\end{equation}
where $b_1^{(4)}=m$, $b_2^{(4)}=b_1^{(4)}+\sigma$, $\chi_1^{(4)}=C$ and $\chi_2^{(4)}=(j+1)\chi_1^{(4)}$.

\subsubsection{Rate-Selective RS}\label{Sec:SSDF}
Dual-hop transmission incurs a pre-log penalty factor of $1/2$. To deal with this rate loss, pure RS is considered only if it provides higher achievable rate than that of the direct $S \longrightarrow D$ transmission \cite{J:Agisilaos_ASIMOLAR_08, J:Win_RS_08}, i.e., higher than $\log_2\left(1+g_{0}\right)$. Using instantaneous CSI and \eqref{Eq:g_end_PureRS}, rate-selective RS chooses between direct (non-relay assisted) and relay-assisted transmission based on the following criterion \cite[eq. (14)]{J:Geo_Agi_RS_2010}
\begin{equation}\label{Eq:SSDF_criterion}
\mathcal{R}_{\text{sel}} = \max\left\{\frac{1}{2}\log_2\left(1+g_{\text{end}}\right),\log_2\left(1+g_{0}\right)\right\}.
\end{equation}
As shown in \cite{J:Geo_Agi_RS_2010}, the MGF of ${g_{\text{sel}}}$ can be obtained using the $M_{g_{\text{end}}}(s)$ of pure RS as
\begin{equation}\label{Eq:MGF_SSDF}
M_{g_{\text{sel}}}(s)=M_{g_0}(s)F_{g_{\text{end}}}\left(\alpha\right) + M_{g_{\text{end}}}(s)[1-F_{g_{\text{end}}}\left(\alpha\right)]
\end{equation}
where $\alpha = g_0^2+2g_0$ is a RV with CDF given by $F_\alpha(x)=F_{g_0}\left(\sqrt{x+1}-1\right)$ which can be obtained using inverse sampling \cite{B:Cochran}. Substituting \eqref{Eq:MGF_end_Final} and \eqref{Eq:OP_FSDF_Difm} to \eqref{Eq:MGF_SSDF}, a closed-form expression for $M_{g_{\text{sel}}}\left(s\right)$ of rate-selective RS over INID Nakagami-$m$ fading with integer $m_\ell$'s and distinct $C_\ell$'s is derived as
\begin{equation}\label{Eq:MGF_Rate_Distinct}
\begin{split}
&M_{g_{\text{sel}}}\left(s\right) =
\frac{C_0^{m_0}}{\left(s+C_0\right)^{m_0}}\left(\prod_{k=1}^L\mathcal{P}_k\right)\left\{\left\{\mathcal{Z}_{m_0}\left(C_0,\alpha\right)
+\sum_{\left\{\lambda_k\right\}_{k=1}^{L}}\sum_{\left\{\mu_j\right\}_{j=1}^{k}}\left(-1\right)^j\left(\prod_{n=1}^k
\frac{1-\mathcal{P}_{\lambda_n}}{\mathcal{P}_{\lambda_n}}\right)\right.\right.
\\&\left.\left.\left.\left\{-C_0^{m_0}\left\{\mathcal{X}_{2,1}^{(1)}\mathcal{Z}_{0}\left[\chi_2^{(1)},\alpha\right]
+\chi_2^{(1)}\mathcal{Y}_1^{(1)}\left(\alpha\right)\right\}+\sum_{\left\{\nu_i\right\}_{i=1}^{j}}\left(\prod_{s=1}^p
\sum_{i_{\nu_s}=1}^{m_{\lambda_{\mu_{\nu_s}}}-1}\frac{C_{\lambda_{\mu_{\nu_s}}}^{i_{\nu_s}}}{i_{\nu_s}!}\right)
C_0^{m_0}\right.\right.\right.\right.
\\&\left.\left.\left.\times\left(\sum_{s=1}^pi_{\nu_s}\right)!\left[\mathcal{Y}_2^{(2)}\left(\alpha\right)-\xi_{p,j}\mathcal{Y}_2^{(3)}\left(\alpha\right)\right]\right\}\right\}\right.
+\left\{1+\sum_{\left\{\lambda_k\right\}_{k=1}^{L}}\sum_{\left\{\mu_j\right\}_{j=1}^{k}}\left(-1\right)^j\left(\prod_{n=1}^k
\frac{1-\mathcal{P}_{\lambda_n}}{\mathcal{P}_{\lambda_n}}\right)\right.
\\&\left.\times\left.\left\{-\left(\sum_{p=1}^jC_{\lambda_{\mu_p}}\right)\left(s+\sum_{p=1}^jC_{\lambda_{\mu_p}}\right)^{-1}
+\sum_{\left\{\nu_i\right\}_{i=1}^{j}}
\left(\sum_{s=1}^pi_{\nu_s}\right)!\left(\prod_{s=1}^p
\sum_{i_{\nu_s}=1}^{m_{\lambda_{\mu_{\nu_s}}}-1}\frac{C_{\lambda_{\mu_{\nu_s}}}^{i_{\nu_s}}}{i_{\nu_s}!}\right)\right.\right.\right.
\\&\left.\left.\left.\times\left(s+\xi_{p,j}\right)^{-\sum_{s=1}^pi_{\nu_s}}\left[1-\xi_{p,j}\left(s+\xi_{p,j}\right)^{-1}\right]\right\}\right.\right.
\\&\left.\left\{1-\left(\prod_{k=1}^L\mathcal{P}_k\right)\left\{\mathcal{Z}_{m_0}\left(C_0,\alpha\right)
+\sum_{\left\{\lambda_k\right\}_{k=1}^{L}}\sum_{\left\{\mu_j\right\}_{j=1}^{k}}\left(-1\right)^j\left(\prod_{n=1}^k
\frac{1-\mathcal{P}_{\lambda_n}}{\mathcal{P}_{\lambda_n}}\right)\right.\right.\right.
\\&\left.\left.\left.\left.\left\{-C_0^{m_0}\left\{\mathcal{X}_{2,1}^{(1)}\mathcal{Z}_{0}\left[\chi_2^{(1)},\alpha\right]
+\chi_2^{(1)}\mathcal{Y}_1^{(1)}\left(\alpha\right)\right\}+\sum_{\left\{\nu_i\right\}_{i=1}^{j}}\left(\prod_{s=1}^p
\sum_{i_{\nu_s}=1}^{m_{\lambda_{\mu_{\nu_s}}}-1}\frac{C_{\lambda_{\mu_{\nu_s}}}^{i_{\nu_s}}}{i_{\nu_s}!}\right)\right.\right.\right.\right.\right.
\\&\left.\left.\left.\left.\times C_0^{m_0}\left(\sum_{s=1}^pi_{\nu_s}\right)!\left[\mathcal{Y}_2^{(2)}\left(\alpha\right)-\xi_{p,j}\mathcal{Y}_2^{(3)}\left(\alpha\right)\right]
\right\}\right\}\right\}\right\}
\end{split}
\end{equation}
while, substituting \eqref{Eq:MGF_end_Final_Equalm} and \eqref{Eq:OP_FSDF_Equalm} to \eqref{Eq:MGF_SSDF}, yields the following closed-form expression for $M_{g_{\text{sel}}}\left(s\right)$ of rate-selective RS over IID Nakagami-$m$ fading channels with integer $m$:
\begin{equation}\label{Eq:MGF_Rate_Equal}
\begin{split}
&\mathcal{M}_{g_{\text{sel}}}\left(s\right) = \frac{C^mL\mathcal{P}^L}{\left(s+C\right)^{m}}
\left\{\left\{\frac{\mathcal{Z}_{m}\left(C,\alpha\right)}{L}+\left(1-\mathcal{P}\right)\mathcal{Z}_{2m}\left(C,\alpha\right)+
\sum_{k=1}^{L-1}\binom{L-1}{k}\frac{\left(1-\mathcal{P}\right)^k}{\mathcal{P}^{k+1}}\right.\right.
\\&\left.\left.\times\left\{\mathcal{P}\mathcal{Z}_{m}\left(C,\alpha\right)+\left(1-\mathcal{P}\right)\mathcal{Z}_{2m}\left(C,\alpha\right)+\sum_{j=1}^k\binom{k}{j}
\left(-1\right)^j\left[\mathcal{P}\mathcal{Z}_{m}\left(C,\alpha\right)\right.\right.\right.\right.
\\&\left.\left.\left.\left.+\sum_{n_1+n_2+\cdots+n_m=j}^j\left(1-\mathcal{P}\right)
\frac{\prod_{i=1}^mn_i^{-1}C^{2m+\sigma}}{\left(m-1\right)!\prod_{j=2}^{m-1}\left(j!\right)^{n_{j+1}}}\mathcal{Y}_2^{(4)}
\left(\alpha\right)\right]\right\}\right\}+\left\{\frac{1}{L}+\left(1-\mathcal{P}\right)C^m\right.\right.
\\&\left.\left.\times\left(s+C\right)^{-m}+\sum_{k=1}^{L-1}\binom{L-1}{k}\frac{\left(1-\mathcal{P}\right)^k}{\mathcal{P}^{k+1}}
\left\{\mathcal{P}+\left(1-\mathcal{P}\right)C^m\left(s+C\right)^{-m}+\sum_{j=1}^k\binom{k}{j}
\right.\right.\right.
\\&\left.\left.\left.\times\left(-1\right)^j\left[\mathcal{P}+\sum_{n_1+n_2+\cdots+n_m=j}^j\frac{\left(1-\mathcal{P}\right)C^{m+\sigma}}{\prod_{j=2}^{m-1}\left(j!\right)^{n_{j+1}}}
\frac{\prod_{i=1}^mn_i^{-1}}{\left(m-1\right)!}\left[s+(j+1)C\right]^{-\left(m+
\sigma\right)}\right]\right\}\right\}\right.
\\&\times\left.\left\{1-L\mathcal{P}^L
\left\{\frac{\mathcal{Z}_{m}\left(C,\alpha\right)}{L}+\left(1-\mathcal{P}\right)\mathcal{Z}_{2m}\left(C,\alpha\right)+
\sum_{k=1}^{L-1}\binom{L-1}{k}\frac{\left(1-\mathcal{P}\right)^k}{\mathcal{P}^{k+1}}
\right.\right.\right.
\\&\left.\left.\left.\times\left\{\mathcal{P}\mathcal{Z}_{m}\left(C,\alpha\right)+\left(1-\mathcal{P}\right)\mathcal{Z}_{2m}\left(C,\alpha\right)+\sum_{j=1}^k\binom{k}{j}
\left(-1\right)^j\left[\mathcal{P}\mathcal{Z}_{m}\left(C,\alpha\right)\right.\right.\right.\right.\right.
\\&\left.\left.\left.\left.\left.+\sum_{n_1+n_2+\cdots+n_m=j}^j\left(1-\mathcal{P}\right)
\frac{\prod_{i=1}^mn_i^{-1}C^{2m+\sigma}}{\left(m-1\right)!\prod_{j=2}^{m-1}\left(j!\right)^{n_{j+1}}}\mathcal{Y}_2^{(4)}
\left(\alpha\right)\right]\right\}\right\}\right\}\right\}.
\end{split}
\end{equation}

\section{Performance Analysis of ODF Relaying Schemes}\label{Sec:Performance}
In this section, the performance of ODF relaying schemes with repetitive and RS-based transmission over INID Nakagami-$m$ fading channels will be analyzed. We will present closed-form and analytical expressions, respectively, for the following performance metrics: \emph{i}) OP and \emph{ii}) ASEP of several modulation formats.

\subsection{Repetitive Transmission}\label{Sec:Performance_Repetitive}
Using the closed-form expressions for $F_{g_{\text{end}}}(x)$ and $M_{g_{\text{end}}}(s)$ presented in Section~\ref{Sec:Repetitive}, the OP and ASEP of repetitive transmission with both MRD and SD are easily obtained as follows.

\emph{OP:} The end-to-end OP of repetitive transmission is easily obtained using $F_{g_{\text{end}}}(x)$ as
\begin{equation}\label{Eq:OP_Rep}
P_{\text{out}} = F_{g_{\text{end}}}\left[2^{(L+1)\mathcal{R}}-1\right].
\end{equation}
Substituting \eqref{Eq:CDF_gEND_MRD_Distinct_C} and \eqref{Eq:CDF_gEND_MRD_Same_C} to \eqref{Eq:OP_Rep}, closed-form expressions for the $P_{\text{out}}$ of repetitive transmission with MRD over INID Nakagami-$m$ fading channels with integer $m_\ell$'s and distinct $C_\ell$'s as well as with arbitrary $m_\ell$'s and $C_\ell=C$ $\forall\,\ell$, respectively, are obtained. Similarly, substituting \eqref{Eq:CDF_gEND_SD_Distinct_C} with \eqref{Eq:Gamma_CDF} and \eqref{Eq:CDF_gbest_SD_Distinct_C} to \eqref{Eq:OP_Rep}, yields a closed-form expression for the $P_{\text{out}}$ of repetitive transmission with SD over INID Nakagami-$m$ fading channels with arbitrary values for $m_\ell$'s.

\emph{ASEP:} Following the MGF-based approach \cite{B:Sim_Alou_Book} and using the $M_{g_{\text{end}}}(s)$ expressions given by \eqref{Eq:MGF_end_Def_Rep} for repetitive transmission with MRD and \eqref{Eq:MGF_end_Final_repSD} and \eqref{Eq:MGF_gend_FINAL_EqualC} for repetitive transmission with SD, the ASEP of several modulation formats for both relaying schemes over INID Nakagami-$m$ fading channels can be easily evaluated. For example, the ASEP of non-coherent binary frequency shift keying (NBFSK) and differential binary phase shift keying (DBPSK) modulation schemes can be directly calculated from $M_{g_{\text{end}}}(s)$; the average bit error rate probability (ABEP) of NBFSK is given by $\overline{P}_{\rm b}=0.5\mathcal{M}_{g_{\text{end}}}(0.5)$ and of DBPSK by $\overline{P}_{\rm b}=0.5\mathcal{M}_{g_{\text{end}}}(1)$. For other schemes, including binary phase shift keying (BPSK), $M$-ary phase shift keying ($M$-PSK)\footnote{It is noted that, for modulation order $M>2$, Gray encoding is assumed, so that $\overline{P}_{\rm s} = \overline{P}_{\rm b}\log_{2}(M)$.}, quadrature amplitude modulation ($M$-QAM), amplitude modulation ($M$-AM), and differential phase shift keying ($M$-DPSK), single integrals with finite limits and integrands composed of elementary functions (exponential and trigonometric) have to be readily evaluated via numerical integration \cite{B:Sim_Alou_Book}. For example, the ASEP of $M$-PSK is easily obtained as
\begin{equation}\label{Eq:MPSK}
\overline{P}_{\rm s}=\frac{1}{\pi}\int_{0}^{\pi-\pi/M} \mathcal{M}_{g_{\text{end}}}\left(\frac{g_{\rm PSK}}{\sin^2\varphi}\right){\rm d}\varphi
\end{equation}
where $g_{\rm PSK}=\sin^{2}\left(\pi/M\right)$, while for $M$-QAM, the ASEP can be evaluated as
\begin{equation}\label{Eq:MQAM}
\begin{split}
\overline{P}_{\rm s}=& \frac{4}{\pi}\left(1-\frac{1}{\sqrt{M}}\right)\left[\int_{0}^{\pi/2} \mathcal{M}_{g_{\text{end}}}\left(\frac{g_{\rm QAM}}{\sin^2\varphi}\right){\rm d}\varphi\right.
\\& \left.-\left(1-\frac{1}{\sqrt{M}}\right)\int_{0}^{\pi/4} \mathcal{M}_{g_{\text{end}}}\left(\frac{g_{\rm QAM}}{\sin^2\varphi}\right){\rm d}\varphi\right]
\end{split}
\end{equation}
with $g_{\rm QAM}=3/\left[2\left(M-1\right)\right]$.

\subsection{RS-based Transmission}\label{Sec:Performance_RS}
The closed-form expressions for the statistics of pure and rate-selective RS presented in Section~\ref{Sec:RS-based} can be easily used to obtain the OP and ASEP of both RS-based schemes as follows.

\emph{OP:} Using the $F_{g_{\text{end}}}(x)$ expressions given by \eqref{Eq:OP_FSDF_Difm} and \eqref{Eq:OP_FSDF_Equalm} over INID and IID Nakagami-$m$ fading channels, respectively, with integer fading parameters, the end-to-end OP of pure RS is easily obtained as
\begin{equation}\label{Eq:OP_FSDF_Def2}
P_{\text{out}} = F_{g_{\text{end}}}\left(2^{2\mathcal{R}}-1\right).
\end{equation}

For rate-selective RS, a closed-form expression for $P_{\text{out}}$ over INID Nakagami-$m$ fading with arbitrary $m_\ell$'s can be easily obtained by substituting  \eqref{Eq:Gamma_CDF} and \eqref{Eq:CDF_gbest_SD_Distinct_C} to \cite[eq. (15)]{J:Geo_Agi_RS_2010}, yielding
\begin{equation}\label{Eq:OP_SSDF}
P_{\text{out}} = \frac{\mathsf{\Gamma}\left[m_0,C_{0}\left(2^{\mathcal{R}}-1\right)\right]}{\Gamma\left(m_0\right)}
\prod_{k=1}^L\left\{\mathcal{P}_ku\left(2^{2\mathcal{R}}-1\right)+\left(1-\mathcal{P}_k\right)
\frac{\mathsf{\Gamma}\left[m_k,C_{k}\left(2^{2\mathcal{R}}-1\right)\right]}{\Gamma\left(m_k\right)}\right\}.
\end{equation}

\emph{ASEP:} Following the MGF-based approach and using the $M_{g_{\text{end}}}(s)$ expressions for pure RS given by \eqref{Eq:MGF_end_Final} and \eqref{Eq:MGF_end_Final_Equalm} over INID and IID Nakagami-$m$ fading channels, respectively, with integer fading parameters, the ASEP of several modulation formats for pure RS can be easily calculated. Similarly, the ASEP for rate-selective RS can be easily evaluated using the $M_{g_{\text{sel}}}(s)$ expressions given by \eqref{Eq:MGF_Rate_Distinct} and \eqref{Eq:MGF_Rate_Equal} over INID and IID Nakagami-$m$ fading channels, respectively, with integer values for the fading parameters.

\section{Numerical Results and Discussions}\label{Sec:num_res}
The analytical expressions of the previous section have been used to evaluate the performance of ODF cooperative systems utilizing repetitive and RS-based transmission over INID Nakagami-$m$ fading channels. Without loss of generality, it has been assumed that $S$'s transmit rate $\mathcal{R}=1$ bps/Hz and that the fading parameters of the first and the second hop of all $L$ links are equal, i.e., $m_k=m_{L+k}$ $\forall\,k$, as well as $\overline{\gamma}_k=\overline{\gamma}_{L+k}$ $\forall\,k$. Moreover, for the performance evaluation results we have considered the exponentially power decaying profile $\overline{\gamma}_k=\overline{\gamma}_0\exp\left(-\ell\delta\right)$ with $\delta$ being the power decaying factor. Clearly, wherever ID fading is considered, $\delta=0$ and $\overline{\gamma}_k=\overline{\gamma}_0=\overline{\gamma}$ $\forall\,k$. In all figures that follow, analytical results for the $P_{\text{out}}$ and $\overline{P}_{\text{b}}$ match perfectly with equivalent performance evaluation results obtained by means of Monte Carlo simulations, thus validating our analysis.

Fig.~\ref{Fig:fig_OP_vs_USERS} illustrates $P_{\text{out}}$ of the considered relaying schemes with both repetitive and RS-based transmission as a function of $L$ for average transmit SNR $\overline{\gamma}_0=5$ dB over IID Nakagami-$m$ fading channels with different values of $m$. As clearly shown, $P_{\text{out}}$ improves with increasing $L$ for both RS-based transmission schemes whereas it degrades severely for both repetitive ones. In particular, utilizing repetitive transmission for $\overline{\gamma}_0=5$ dB is unavailing for $L\geq3$ irrespective of the fading conditions. As for RS-based transmission, the gains from RS diminish as $m$ increases; the smaller the $m$, the greater the $P_{\text{out}}$ gain from relaying. Furthermore, pure RS is inefficient when $L$ is small and rate-selective RS leads to significant $P_{\text{out}}$ gains over the pure one as $m$ decreases. The impact of increasing $L$ in the $P_{\text{out}}$ performance of all four relaying schemes is also demonstrated in Fig.~\ref{Fig:fig_OP_vs_USERS_difSNRs} for different values of $\overline{\gamma}_0$ and IID Rayleigh fading conditions. As shown from this figure and Fig.~\ref{Fig:fig_OP_vs_USERS} for both repetitive and RS-based transmission, $P_{\text{out}}$ degrades with decreasing $\overline{\gamma}_0$ and/or $m$. For example, although for $\overline{\gamma}_0=5$ dB repetitive transmission does not benefit from relaying for $L\geq3$, this happens for $\overline{\gamma}_0=0$ dB when $L\geq2$. Moreover, for the latter transmit SNR, the improvement on the $P_{\text{out}}$ of pure RS with increasing $L$ is very small, while $P_{\text{out}}$ performance of rate-selective RS is irrespective of $L$; the smaller the $\overline{\gamma}_0$, the smaller the gains from relaying. This happens because as fading conditions become more severe and transmit SNR reduces, the gains from RS eventually decrease.

Using \eqref{Eq:Prob_k} in Fig.~\ref{Fig:fig_Decod_Prob}, the decoding probability $1-\mathcal{P}_k$ is plotted versus $L$ for both repetitive and RS-based transmission over IID Nakagami-$m$ fading conditions for the source to relay channels with different values of $m$ and $\overline{\gamma}_{L+k}=\overline{\gamma}$ $\forall\,k,L$. As expected, decreasing $m$ and/or $\overline{\gamma}$ decreases the decoding probability. Moreover, this probability is severely degraded with increasing $L$ for repetitive transmission, whereas it remains unchanged with increasing $L$ for the RS-based one. The $P_{\text{out}}$ performance of pure and rate-selective RS is depicted in Fig.~\ref{Fig:fig_OP_vs_dB_RS} versus $\overline{\gamma}_0$ for $L=2$ relays over INID Nakagami-$m$ fading channels with different values of $m_0$, $m_1$ and $m_2$. As expected, for both RS-based schemes, $P_{\text{out}}$ improves with increasing $\overline{\gamma}_0$ and/or any of the fading parameters. More importantly, it is shown that, as the fading conditions of the relay to destination channels become more favorable than those of the direct source to destination channel, RS-based transmission improves $P_{\text{out}}$. On the contrary, whenever the fading conditions of the relay to destination channels are similar to the direct one, non-relay assisted transmission results in lower $P_{\text{out}}$ than that of the RS-based one.

In Fig.~\ref{Fig:fig_BER_vs_USERS}, the $\overline{P}_{\text{b}}$ performance of square $4$-QAM is plotted as a function of $L$ for average SNR per bit $\overline{\gamma}_{\text{b}}=5$ dB for all second hop channels and $\overline{\gamma}_0=0$ dB over IID Nakagami-$m$ fading channels with different values of $m$. It is clearly shown that the $\overline{P}_{\text{b}}$ of RS-based transmission improves with increasing $L$ and/or $m$ whereas for both repetitive schemes, although $\overline{P}_{\text{b}}$ improves with increasing $m$, it does not benefit from increasing $L$. Interestingly, for RS-based transmission, as $L$ and/or $m$ increase, the $\overline{P}_{\text{b}}$ improvement of pure compared with rate-selective RS increases; as shown in \eqref{Eq:SSDF_criterion}, rate-selective RS might choose the direct transmission even in cases where the received SNR from the best relay is larger that from the source node. Assuming IID Rayleigh fading conditions, Fig.~\ref{Fig:fig_BER_vs_USERS_difSNRs} illustrates $\overline{P}_{\text{b}}$ of DBPSK versus $L$ for all considered relaying schemes and for different values of $\overline{\gamma}_0$. As shown from this figure and Fig.~\ref{Fig:fig_BER_vs_USERS} for both repetitive and RS-based transmission, $\overline{P}_{\text{b}}$ for both modulations improves with increasing $\overline{\gamma}_0$ and/or $m$. It was shown in Fig.~\ref{Fig:fig_Decod_Prob} that more favorable fading conditions and larger $\overline{\gamma}_0$ increase the decoding probability, thus both repetitive transmission schemes become unavailing for larger $L$, whereas the gains from RS increase. The $\overline{P}_{\text{b}}$ performance of DBPSK of pure and rate-selective RS is depicted in Fig.~\ref{Fig:fig_BER_vs_dB_RS} versus $\overline{\gamma}_{\text{b}}$ for all second hop channels with $L=1$, $2$ and $3$ relays and for $\overline{\gamma}_0=0$ dB over INID Nakagami-$m$ fading channels with different values for the fading parameters. Clearly, for both RS-based transmission schemes, $\overline{P}_{\text{b}}$ improves with increasing $\overline{\gamma}_{\text{b}}$ and/or $m$ and/or $L$. It is shown that, as $\overline{\gamma}_{\text{b}}$ increases and as the fading conditions of the relay to destination channels become more favorable than those of the direct source to destination channel, RS-based transmission results in larger $\overline{P}_{\text{b}}$ improvement. More importantly, even for strong source to destination channel conditions, as $\overline{\gamma}_{\text{b}}$ increases the improvement on $\overline{P}_{\text{b}}$ with RS-based transmission becomes larger than that with the non-relay assisted one.

\section{Conclusion}\label{Sec:concl}
Cooperative diversity is a very promising avenue for future wireless communications. However, it is necessary to investigate how relays should be utilized in order to achieve certain objectives. In this paper, we presented a general analytical framework for modelling and evaluating performance of four DF relaying schemes under INID Nakagami-$m$ fading channels. Moreover, we obtained closed-form expressions for the OP and the ASEP performance of the RS and repetitive schemes when MRD or SD are employed at the destination node. We concluded that RS performs better than repetitive transmission when CSI for RS is perfect. Furthermore, it was shown that relaying should be switched-off when the source to destination direct link is sufficiently strong and the aim is minimizing OP. However, in terms of ASEP, relaying was shown to be always beneficial.

\bibliographystyle{IEEEtran}
\bibliography{IEEEabrv,Alexandropoulos}

\newpage
\textbf{FIGURES' CAPTIONS}

\begin{figure}[!t]
\centering
\includegraphics[keepaspectratio,width=\columnwidth]{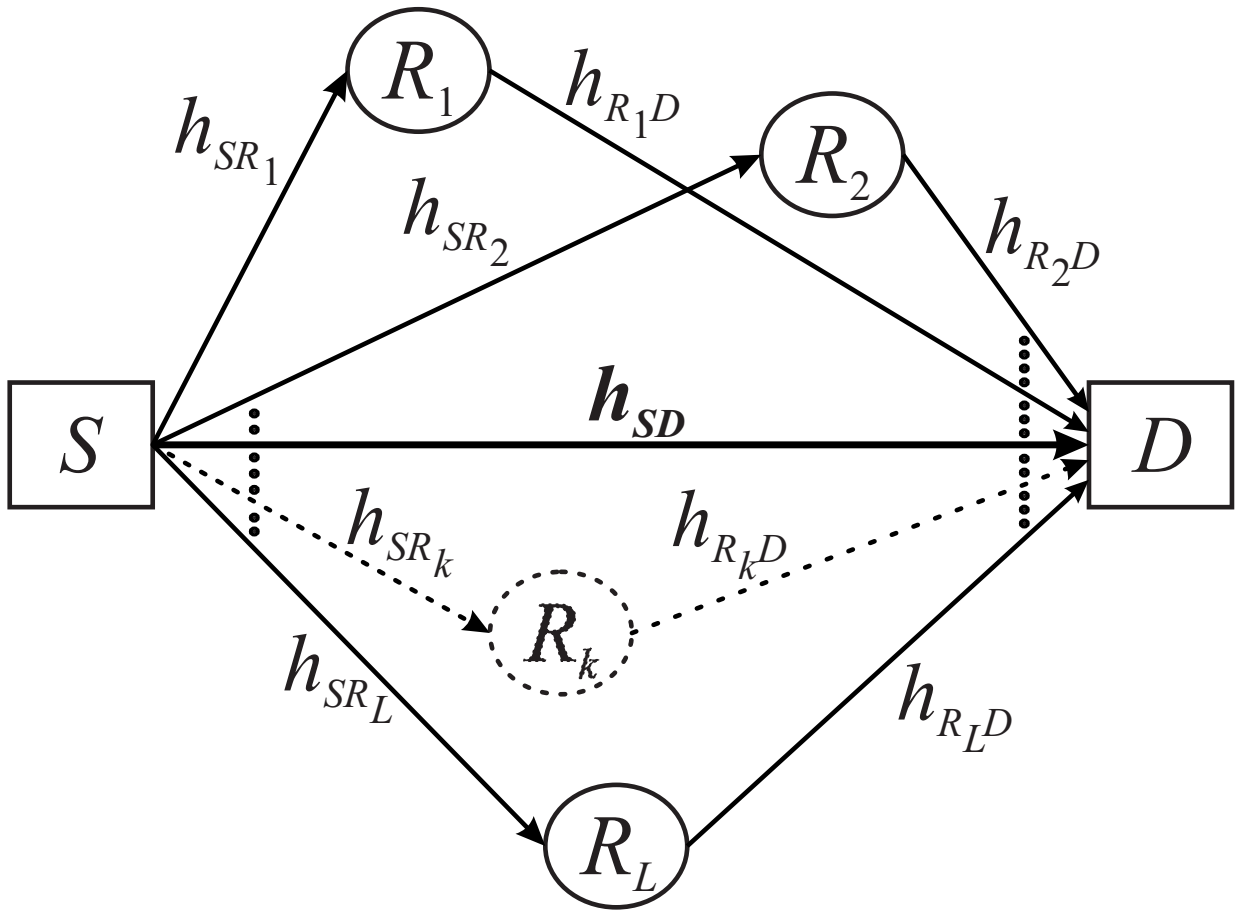}
\caption{Illustration of a dual-hop cooperative wireless system with $L+2$ wireless nodes: a source node $S$, $L$ relay station nodes $R_k$, $k=1,2,\ldots,L$, and a destination node $D$.} \label{Fig:System}
\end{figure}

Fig.~\ref{Fig:System}: Illustration of a dual-hop cooperative wireless system with $L+2$ wireless nodes: a source node $S$, $L$ relay station nodes $R_k$, $k=1,2,\ldots,L$, and a destination node $D$.\\

\begin{figure}[!t]
\centering
\includegraphics[keepaspectratio,width=\columnwidth]{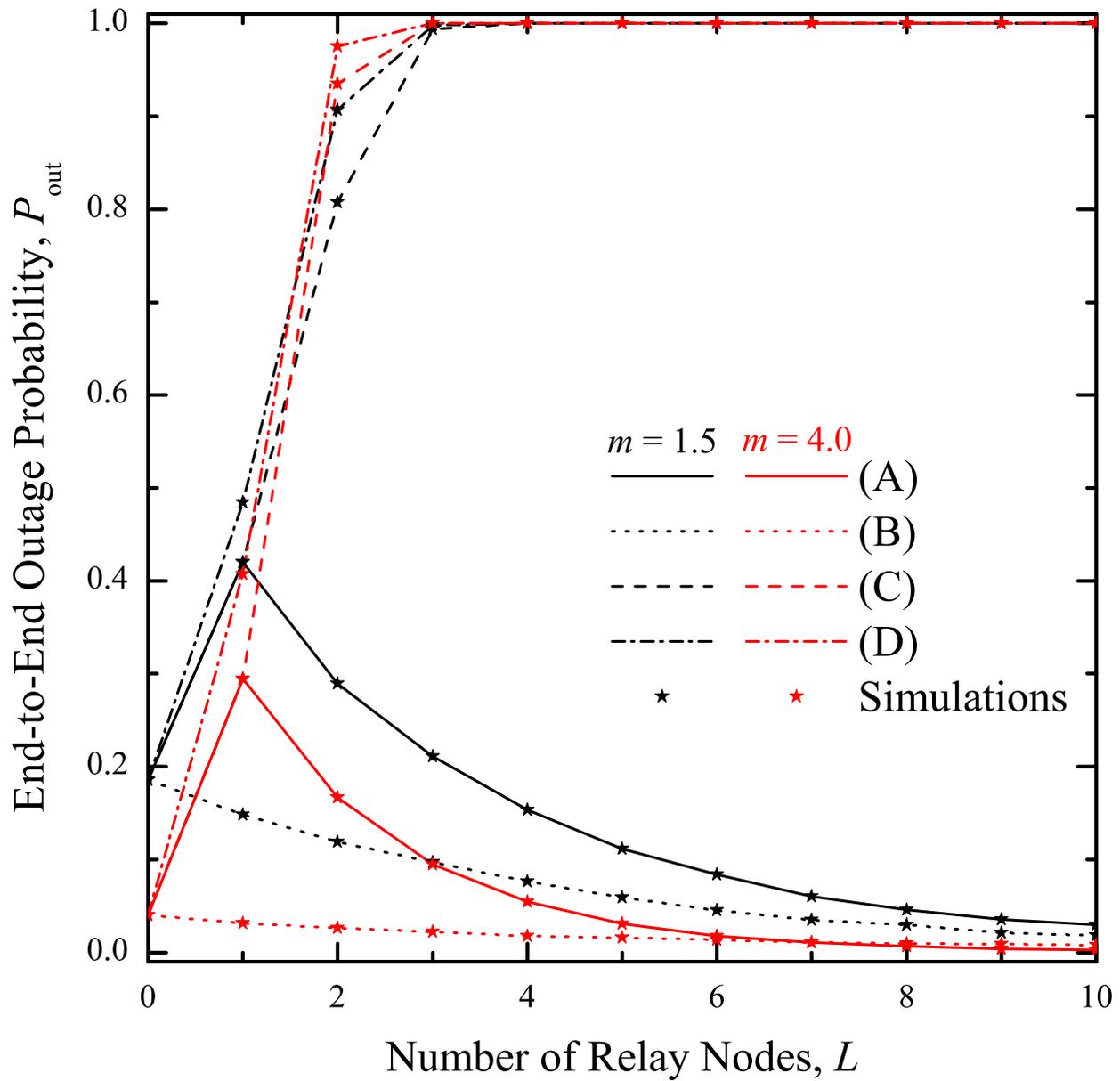}
\caption{End-to-end OP, $P_{\text{out}}$, versus the number of relay nodes, $L$, for average transmit SNR $\overline{\gamma}_0=5$ dB over IID Nakagami-$m$ fading channels with different values of $m$: (A) Pure RS, (B) Rate-Selective RS, (C) Repetitive transmission with MRD and (D) Repetitive transmission with SD.} \label{Fig:fig_OP_vs_USERS}
\end{figure}

Fig.~\ref{Fig:fig_OP_vs_USERS}: End-to-end OP, $P_{\text{out}}$, versus the number of relay nodes, $L$, for average transmit SNR $\overline{\gamma}_0=5$ dB over IID Nakagami-$m$ fading channels with different values of $m$: (A) Pure RS, (B) Rate-Selective RS, (C) Repetitive transmission with MRD and (D) Repetitive transmission with SD.\\

\begin{figure}[!t]
\centering
\includegraphics[keepaspectratio,width=\columnwidth]{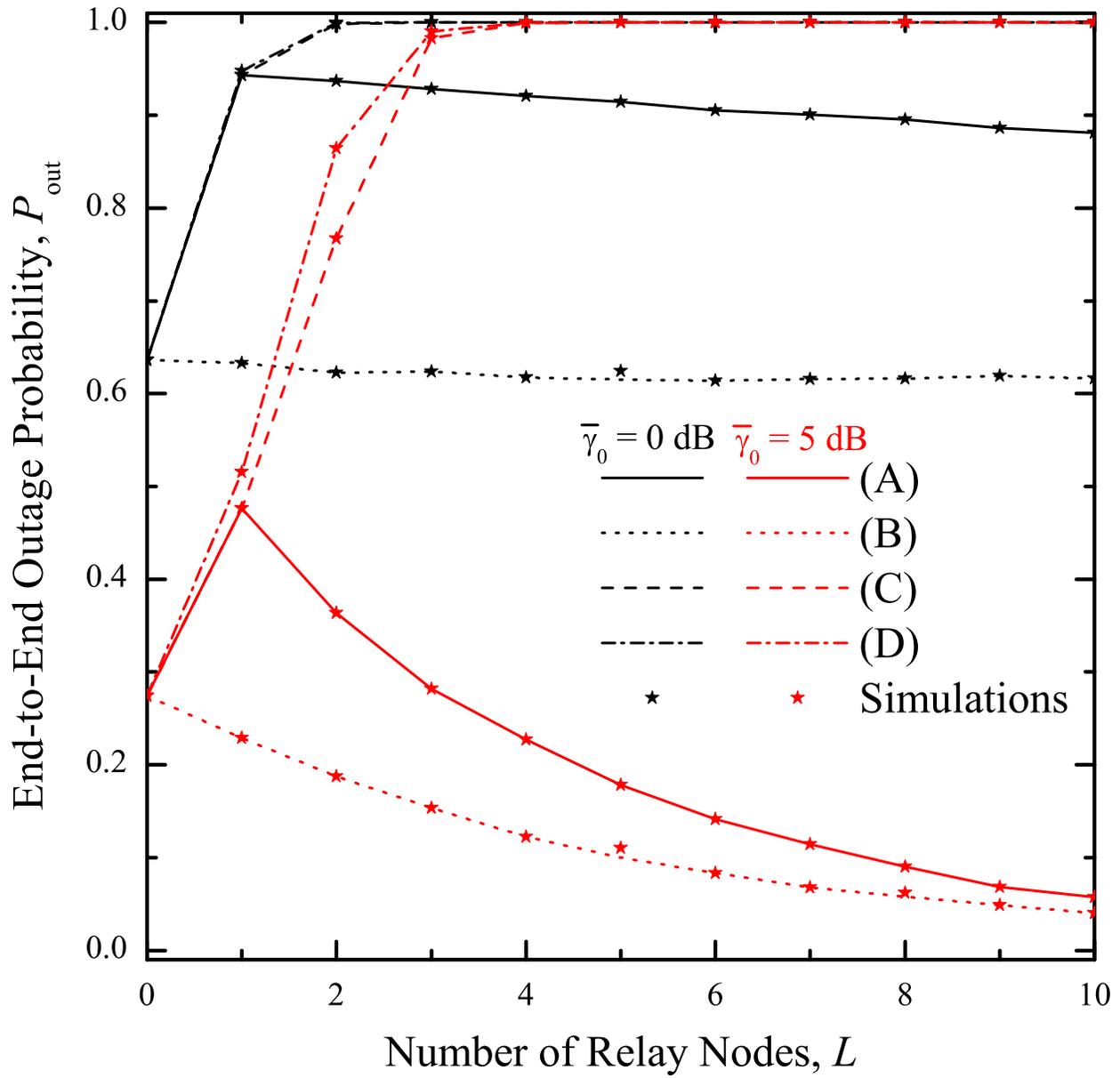}
\caption{End-to-end OP, $P_{\text{out}}$, versus the number of relay nodes, $L$, for different average transmit SNRs over IID Rayleigh fading channels: (A) Pure RS, (B) Rate-Selective RS, (C) Repetitive transmission with MRD and (D) Repetitive transmission with SD.} \label{Fig:fig_OP_vs_USERS_difSNRs}
\end{figure}

Fig.~\ref{Fig:fig_OP_vs_USERS_difSNRs}: End-to-end OP, $P_{\text{out}}$, versus the number of relay nodes, $L$, for different average transmit SNRs over IID Rayleigh fading channels: (A) Pure RS, (B) Rate-Selective RS, (C) Repetitive transmission with MRD and (D) Repetitive transmission with SD.\\

\begin{figure}[!t]
\centering
\includegraphics[keepaspectratio,width=\columnwidth]{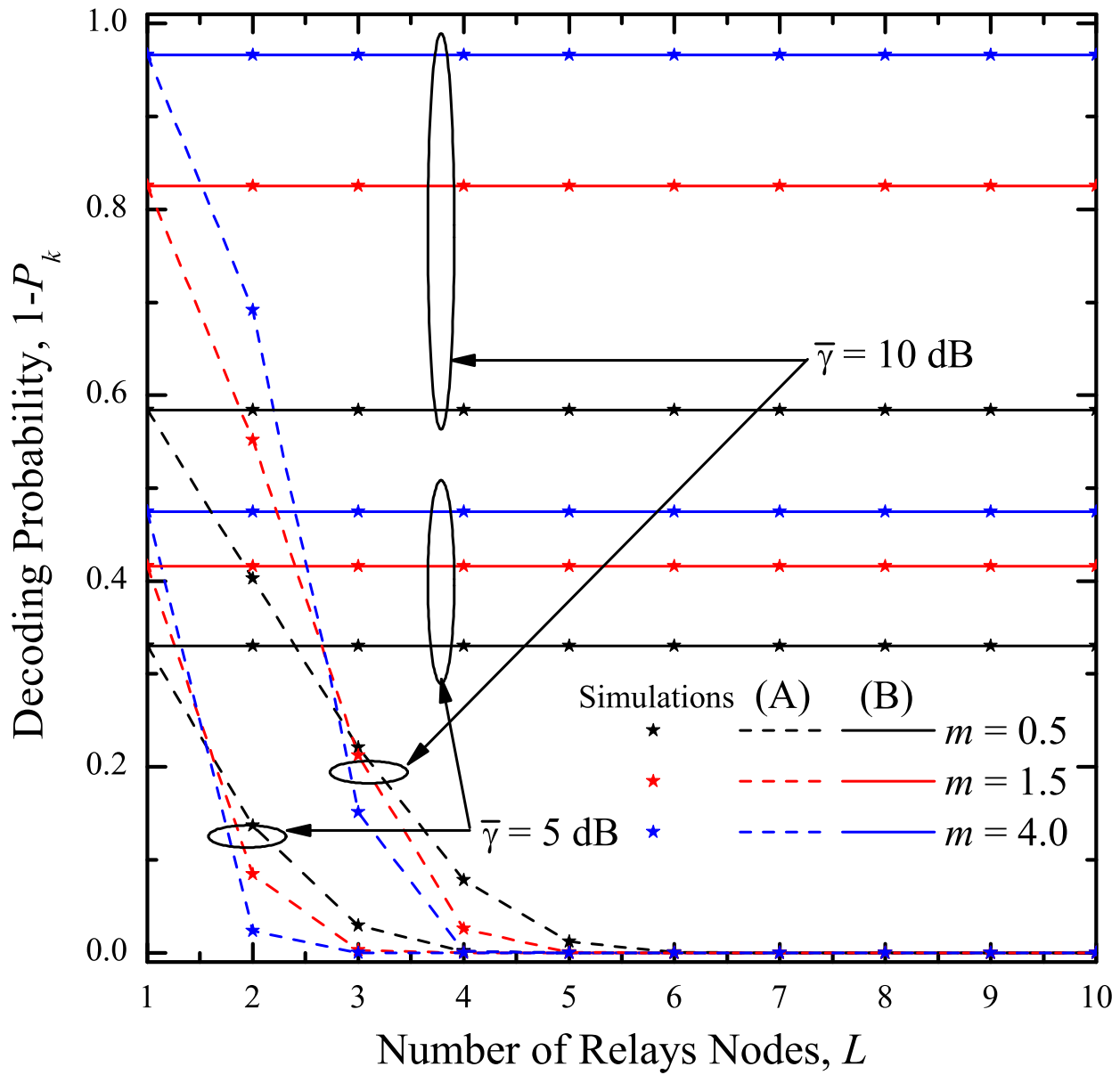}
\caption{Decoding probability, $1-\mathcal{P}_k$, versus the number of relay nodes, $L$, over IID Nakagami-$m$ fading channels with different values of $m$ and $\overline{\gamma}_{L+k}=\overline{\gamma}$ $\forall\,k,L$: (A) Repetitive and (B) RS-based transmission.} \label{Fig:fig_Decod_Prob}
\end{figure}

Fig.~\ref{Fig:fig_Decod_Prob}: Decoding probability, $1-\mathcal{P}_k$, versus the number of relay nodes, $L$, over IID Nakagami-$m$ fading channels with different values of $m$ and $\overline{\gamma}_{L+k}=\overline{\gamma}$ $\forall\,k,L$: (A) Repetitive and (B) RS-based transmission.\\

\begin{figure}[!t]
\centering
\includegraphics[keepaspectratio,width=\columnwidth]{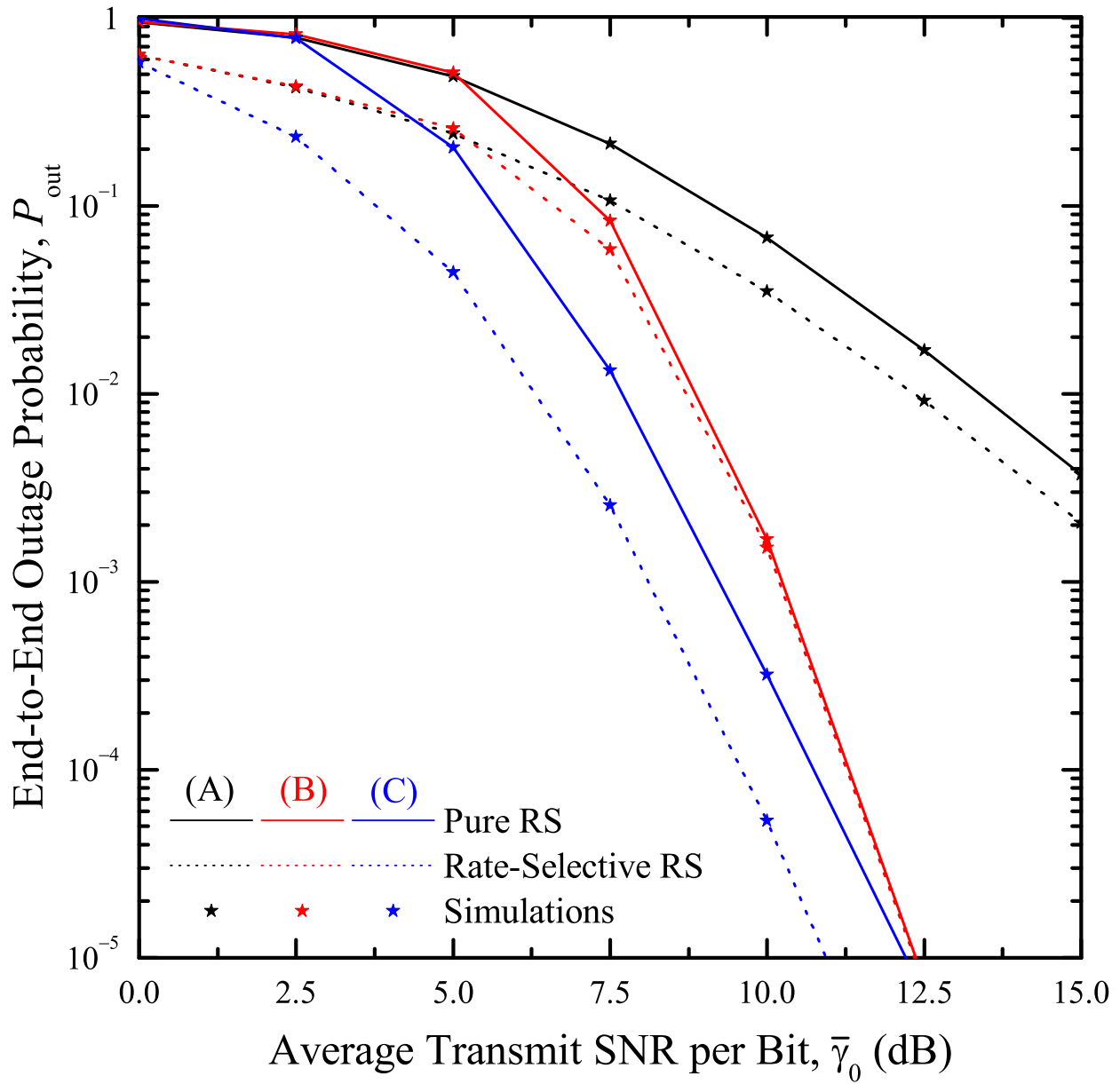}
\caption{End-to-end OP, $P_{\text{out}}$, of RS versus the average transmit SNR per bit, $\overline{\gamma}_0$, for $L=2$ relay nodes over INID Nakagami-$m$ fading channels: (A) $m_0=m_1=m_2=1$ and $\delta=0.3$, (B) $m_0=1$, $m_1=m_2=6$ and $\delta=0.3$ and (C) $m_0=m_1=m_2=3$ and $\delta=0$.} \label{Fig:fig_OP_vs_dB_RS}
\end{figure}

Fig.~\ref{Fig:fig_OP_vs_dB_RS}: End-to-end OP, $P_{\text{out}}$, of RS versus the average transmit SNR per bit, $\overline{\gamma}_0$, for $L=2$ relay nodes over INID Nakagami-$m$ fading channels: (A) $m_0=m_1=m_2=1$ and $\delta=0.3$, (B) $m_0=1$, $m_1=m_2=6$ and $\delta=0.3$ and (C) $m_0=m_1=m_2=3$ and $\delta=0$.\\

\begin{figure}[!t]
\centering
\includegraphics[keepaspectratio,width=\columnwidth]{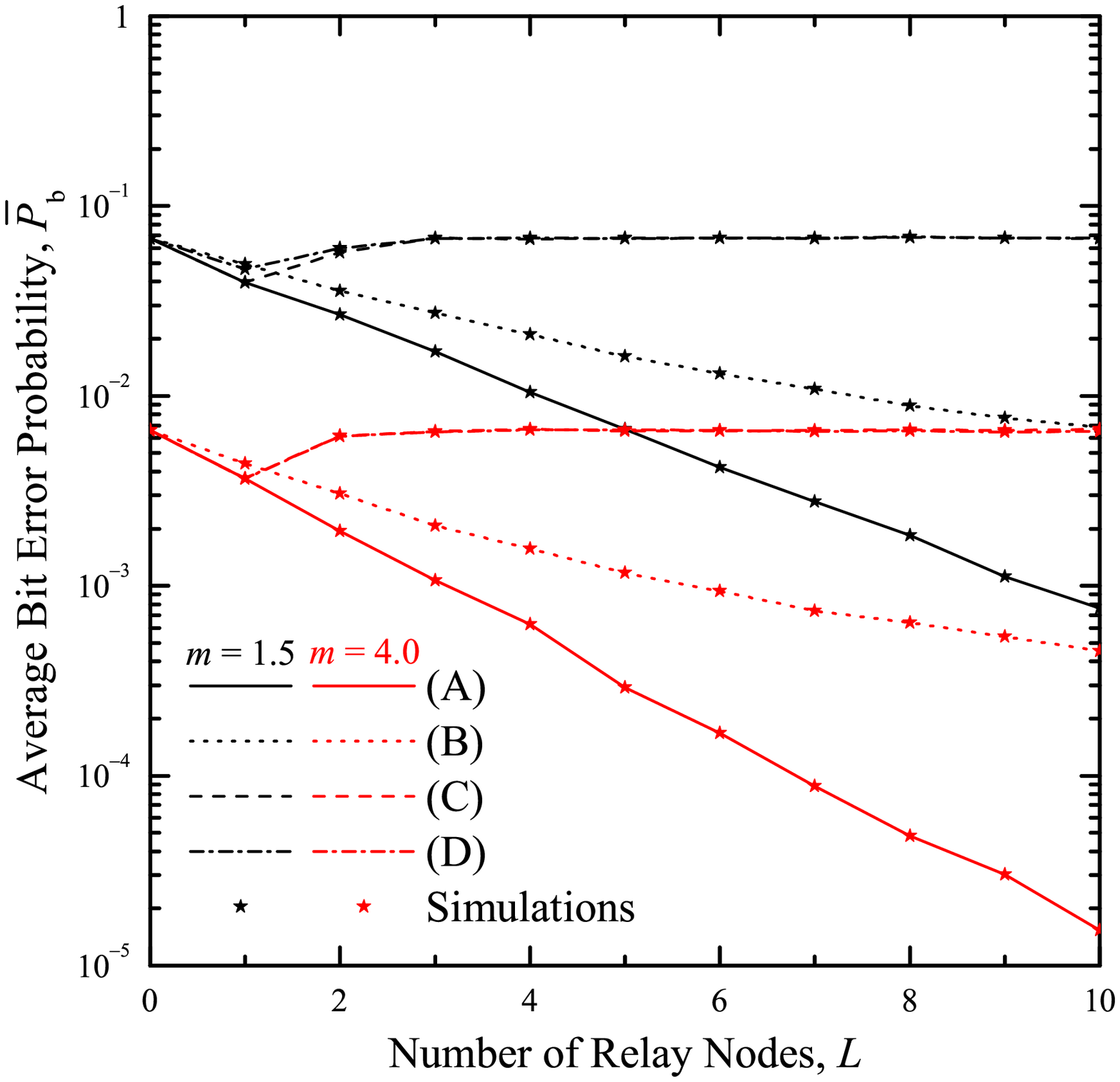}
\caption{ABEP, $\overline{P}_{\text{b}}$, of square $4$-QAM versus the number of relay nodes, $L$, for average transmit SNR $\overline{\gamma}_{\text{b}}=5$ dB and $\overline{\gamma}_0=0$ dB over IID Nakagami-$m$ fading channels with different values of $m$: (A) Pure RS, (B) Rate-Selective RS, (C) Repetitive transmission with MRD and (D) Repetitive transmission with SD.} \label{Fig:fig_BER_vs_USERS}
\end{figure}

Fig.~\ref{Fig:fig_BER_vs_USERS}: ABEP, $\overline{P}_{\text{b}}$, of square $4$-QAM versus the number of relay nodes, $L$, for average transmit SNR $\overline{\gamma}_{\text{b}}=5$ dB and $\overline{\gamma}_0=0$ dB over IID Nakagami-$m$ fading channels with different values of $m$: (A) Pure RS, (B) Rate-Selective RS, (C) Repetitive transmission with MRD and (D) Repetitive transmission with SD.\\

\begin{figure}[!t]
\centering
\includegraphics[keepaspectratio,width=\columnwidth]{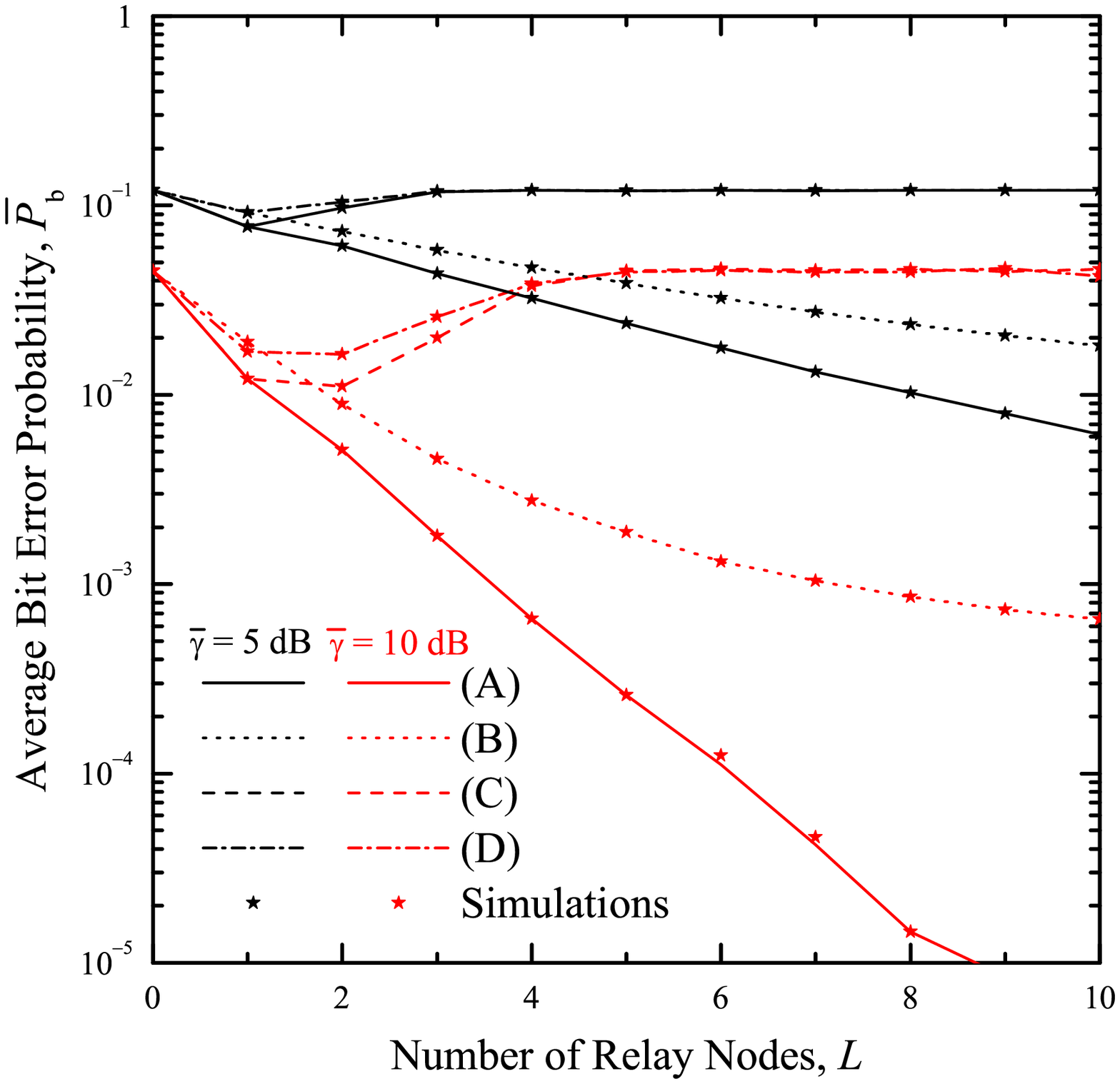}
\caption{ABEP, $\overline{P}_{\text{b}}$, of DBPSK versus the number of relay nodes, $L$, for different average transmit SNRs per bit and $\overline{\gamma}_0=0$ dB, over IID Rayleigh fading channels: (A) Pure RS, (B) Rate-Selective RS, (C) Repetitive transmission with MRD and (D) Repetitive transmission with SD.} \label{Fig:fig_BER_vs_USERS_difSNRs}
\end{figure}

Fig.~\ref{Fig:fig_BER_vs_USERS_difSNRs}: ABEP, $\overline{P}_{\text{b}}$, of DBPSK versus the number of relay nodes, $L$, for different average transmit SNRs per bit and $\overline{\gamma}_0=0$ dB, over IID Rayleigh fading channels: (A) Pure RS, (B) Rate-Selective RS, (C) Repetitive transmission with MRD and (D) Repetitive transmission with SD.\\

\begin{figure}[!t]
\centering
\includegraphics[keepaspectratio,width=\columnwidth]{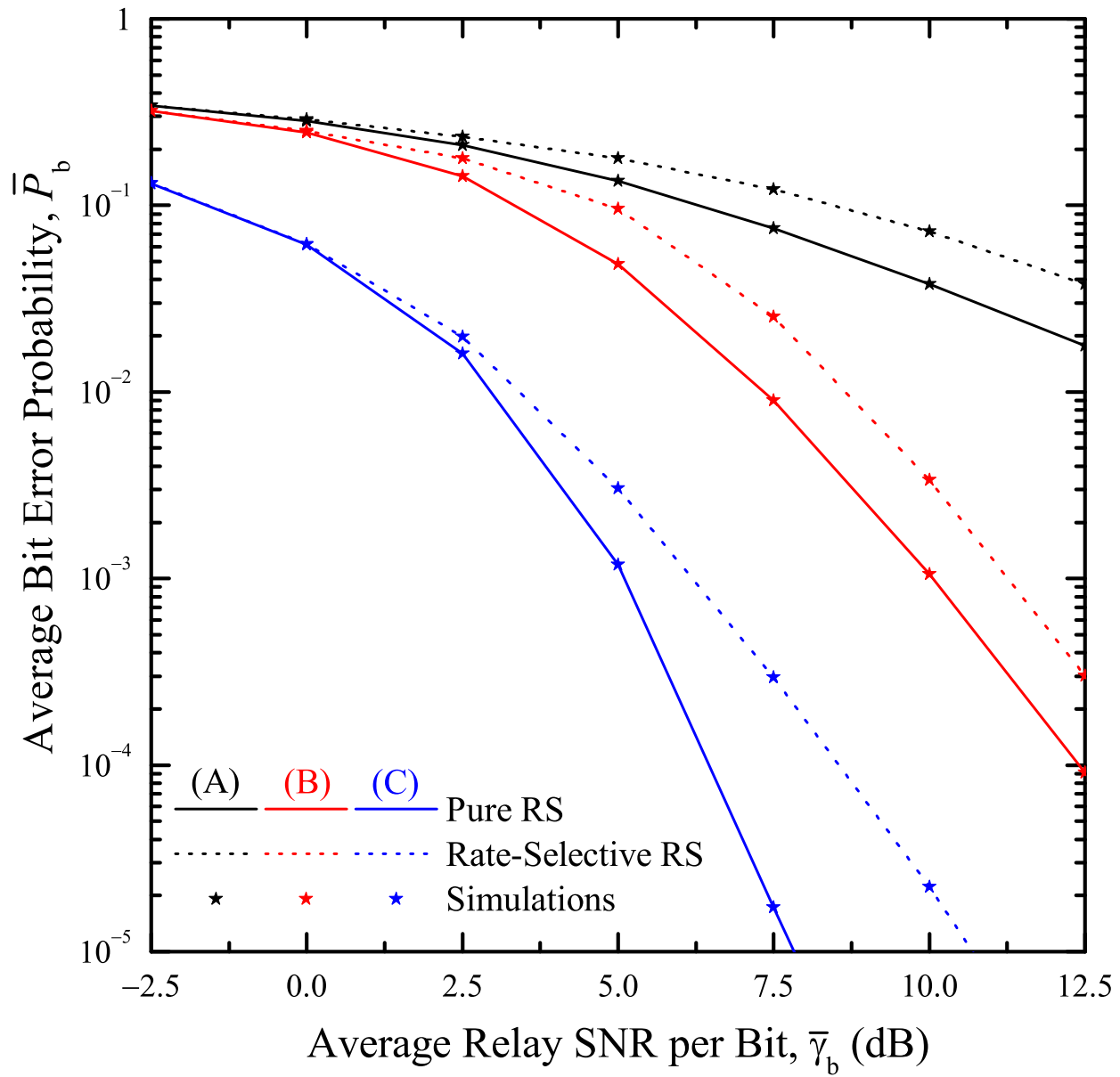}
\caption{ABEP, $\overline{P}_{\text{b}}$, of DBPSK for RS versus the average relay SNR per bit, $\overline{\gamma}_{\text{b}}$, and for $\overline{\gamma}_0=0$ dB over INID Nakagami-$m$ fading channels: (A) $L=1$, $m_0=0.5$, $m_1=1$ and $\delta=0.1$, (B) $L=2$, $m_0=1$, $m_1=m_2=2$ and $\delta=0$ and (C) $L=3$, $m_\ell=3$ for $\ell=0,1,2$ and $3$ and $\delta=0$.} \label{Fig:fig_BER_vs_dB_RS}
\end{figure}

Fig.~\ref{Fig:fig_BER_vs_dB_RS}: ABEP, $\overline{P}_{\text{b}}$, of DBPSK for RS versus the average relay SNR per bit, $\overline{\gamma}_{\text{b}}$, and for $\overline{\gamma}_0=0$ dB over INID Nakagami-$m$ fading channels: (A) $L=1$, $m_0=0.5$, $m_1=1$ and $\delta=0.1$, (B) $L=2$, $m_0=1$, $m_1=m_2=2$ and $\delta=0$ and (C) $L=3$, $m_\ell=3$ for $\ell=0,1,2$ and $3$ and $\delta=0$.\\

\newpage

\newpage

\newpage

\newpage

\newpage

\newpage

\newpage

\newpage

\end{document}